\documentclass{article}
\usepackage[utf8]{inputenc}
\usepackage[left=3cm,right=3cm,top=3cm,bottom=3cm]{geometry}
\usepackage{amsmath}
\usepackage{amssymb}
\usepackage{graphicx}
\usepackage{esint}
\usepackage{enumitem}
\usepackage{multirow}
\usepackage{graphicx}
\usepackage{hhline}
\usepackage{hyperref}
\hypersetup{
  colorlinks   = true,    % Colours links instead of ugly boxes
  urlcolor     = blue,    % Colour for external hyperlinks
  linkcolor    = blue,    % Colour of internal links
  citecolor    = black     % Colour of citations
}

\newcommand\Tr{\mathrm{Tr}}

\usepackage{color}
\usepackage[makeroom]{cancel}
\usepackage[normalem]{ulem}
\definecolor{pkcolor}{rgb}{0,0.1,0.7}

\newcommand\pkout{\marginpar{\color{pkcolor}$\clubsuit$}\bgroup\markoverwith{\color{pkcolor}{\rule[0.4ex]{2pt}{0.8pt}}}\ULon}

\definecolor{hkcolor}{rgb}{0.7,0.0,0.0}

\newcommand\hkout{\marginpar{\color{hkcolor}$\clubsuit$}\bgroup\markoverwith{\color{hkcolor}{\rule[0.4ex]{2pt}{0.8pt}}}\ULon}

\definecolor{ascolor}{rgb}{0.0,0.7,0.3}

\DeclareRobustCommand\asout{\bgroup\markoverwith{\color{ascolor}{\rule[0.4ex]{2pt}{0.8pt}}}\ULon}

\title{A new Wilson line-based action for gluodynamics}
\author{Hiren Kakkad$^a$,
Piotr Kotko$^a$,
Anna Stasto$^b$
\\ \\
$^a${\it AGH University Of Science and Technology, Physics Faculty,} \\ 
{\it Mickiewicza 30, 30-059 Krakow, Poland} \\ \\
$^b${\it The Pennsylvania State University, Physics Department}\\ 
{\it 104 Davey Lab, University Park, PA 16802, USA }
}

\date{}

\begin{document}
%=============================================
\maketitle

\begin{abstract}
    We perform a canonical transformation of fields that brings the Yang-Mills action in the light-cone gauge to a new classical action, which does not involve any triple-gluon vertices. The lowest order vertex is the four-point MHV vertex. Higher point vertices include the MHV and $\overline{\textrm{MHV}}$ vertices, that reduce to the corresponding amplitudes in the on-shell limit. In general, any $n$-leg vertex has $2\leq m \leq n-2$ 
    negative
    helicity legs. The canonical transformation of fields can be compactly expressed in terms of path-ordered exponentials of fields and their functional derivative. We apply the new action to compute several tree-level amplitudes, up to 8-point NNMHV amplitude, and find agreement with the standard methods.
     The absence of triple-gluon vertices results in fewer diagrams required to compute amplitudes, when compared to the CSW method and, obviously, considerably  fewer than in the standard Yang-Mills action.
\end{abstract}

%---------------------------------------------
%---------------------------------------------
\section{Introduction}
\label{sec:Intro}

Although quark and gluon fields are considered to be the most fundamental degrees of freedom of Quantum Chromodynamics (QCD),
 in a variety of situations they are definitely not the most effective ones.
There are several quite distinct aspects of the QCD theory where this becomes manifest.
For example, in QCD factorization theorems, which are
 essential tools to relate the theory to high energy experiments, one needs to account not just for a single quark or gluon exchanged between subprocesses separated by a large energy scale, but also for all collinear and soft gluons;
 these resummed exchanges of gluons can be  included by using   Wilson lines (see \cite{Collins:2011zzd} for a comprehensive review of the factorization theorems).
Similar collective degrees of freedom appear naturally in the high energy limit of QCD (see e.g. \cite{Lipatov:1995pn,Gelis2010}).
 In the domain of  non-perturbative QCD, further example can be provided by  lattice QCD, where instead of the gauge fields one 
uses Wilson lines and Wilson loops. It is also known that, in the limit of large number of colors, the gauge theory can be 
completely formulated in  loop space, i.e. in terms of various contours of the Wilson loop \cite{Polyakov1980} (see also a textbook \cite{Cherednikov2014}).

The subject of interest of the following work are  scattering amplitudes, in particular the pure gluonic amplitudes. 
In that context, it is already well understood that the elementary triple and four-gluon interactions are not the 
most effective bricks to build the amplitudes. Pictorially, they are way too small, so that the number of Feynman 
diagrams can become  intractable for multi gluon processes.
Instead, one should rather use the smaller amplitudes (i.e. with fewer legs) as the building blocks, but they must be deformed to the non-physical domain according to the Britto-Cachazo-Feng-Witten (BCFW) method \cite{Britto:2004ap,Britto:2005fq}. A particularly interesting example of the BCFW method is when the only type of amplitudes used are the maximally helicity violating (MHV) amplitudes \cite{Risager2005}. This type of recursion has been found earlier based on the twistor space formulation of  quantum field theory by Cachazo, Svrcek and Witten (CSW) \cite{Cachazo2004}, who suggested that the MHV amplitudes continued off-shell are really the multileg \emph{vertices} (this formalism was later proved to be equivalent to the standard Yang-Mills theory in \cite{Gorsky_2006}). Indeed, an explicit action has been found, where the MHV vertices are reproduced as a result of the canonical field transformation on the Yang-Mills action in light-cone gauge \cite{Mansfield2006,Ettle2006b}
This so-called "MHV action"
has also been 
developed for spontaneously broken gauge theory in \cite{Buchta_2010} and also for supersymmetric gauge theories in \cite{Morris_2008, Feng_2009, Fu_2010}.
It was also shown that the action reproduces correctly the one-loop same helicity amplitudes  \cite{Ettle2007,Brandhuber2007a,Brandhuber2007,Ettle2008,Feng2009,Boels_2008, Elvang_2012}. Furthermore, in the supersymmetric regime, using the MHV vertices, MHV loop amplitudes in $\mathcal{N
}=4$ super Yang-Mills theory were derived explicitly in \cite{Brandhuber_2005}. This was later extended for $\mathcal{N
}=1$ (and $\mathcal{N
}=2$) super Yang-Mills in \cite{Bedford_2005}. Following this, non-supersymmetric one-loop amplitudes, using MHV diagrams, were explored in \cite{Bedford_2005b}. 

Interestingly, 
the new fields appearing in the MHV action turn out to be related to the straight infinite Wilson line of the Yang-Mills fields \cite{Kotko2017,Kakkad2020}.

As the above result is central to the present paper, let us describe it in more  detail. The light-cone Yang-Mills action can be expressed in terms of just two transverse gluon fields, that correspond  to two polarization states in the on-shell limit. The MHV action is obtained by transforming both  fields to a new pair of fields. In \cite{Kotko2017} it was found that the plus helicity field in the MHV action is given as the straight infinite Wilson line along the complex direction determined by the plus helicity polarization vector. This means that the line lies on the so-called self-dual plane, i.e. the plane on which the tensors are self-dual. In the recent paper \cite{Kakkad2020},
we
found that the minus helicity field is  given by a similar Wilson line, but with an insertion of the minus helicty gluon field somewhere on the line.  Additionally, 
we postulated, that it should be a part of a bigger structure, extending beyond the self-dual plane.

Indeed, in the present work we find a more general canonical transformation based on path ordered exponentials of the gauge fields, extending over both the self-dual and anti-self-dual planes. The field transformation can be most easily derived as a subsequent canonical transformation of the anti-self-dual part of the MHV action, but we also discuss a direct link between the new action and the Yang-Mills action. 
The key property of the new action is that it does not have the triple-gluon vertices at all. The reason for this structure is that the triple-gluon vertices have been effectively resummed inside the Wilson lines. The absence of the $(++-)$ vertex already occurs at the level of the MHV action, as previously demonstrated in \cite{Mansfield2006}. The second canonical transformation of the  anti-self-dual part  of the MHV action results 
in the absence of  $(--+)$  triple-gluon vertex as well. 
Thus, the lowest multiplicity vertex is the four-point MHV vertex.
Higher-point vertices include not only the MHV vertices, but also other helicity configurations. The number of diagrams needed to obtain amplitudes beyond the MHV level is thus greatly reduced.
We perform explicit calculations within the new formulation of several higher multiplicity amplitudes, to verify the consistency of the results.

The paper is organized as follows. In Section~\ref{sec:Z-theory} we introduce the new action, first on a general ground, and then we proceed to a more technical derivation. In Section~\ref{sec:Applications} we apply the new theory to actual amplitude computations. In Section~\ref{sec:Summary} we summarize the work and discuss some aspects of the new action, in particular the geometric picture behind the field transformation. Finally, in the Appendix we provide more technical details of selected calculations.

%---------------------------------------------
%---------------------------------------------
\section{A new classical action for gluodynamics  }
\label{sec:Z-theory}

%-----------------------------------------------
\subsection{General idea}

Motivated by our earlier results \cite{Kotko2017,Kakkad2020}, we look for a new set of classical fields describing scattering processes with gluons in the simplest possible way. That is, we want to find an action, which has interaction vertices as close to the real scattering processes, as possible. It is known, that the lowest non-zero scattering amplitude in the physical domain (i.e. on-shell real momenta satisfying the momentum conservation) is the four-point amplitude. Therefore, we will look for an action that has no triple-field 
coupling.
In addition, we want the new fields to have a closed form in terms of the ordinary Yang-Mills fields. This is 
necessary for practical applications of the new action.

In general, a field transformation 
relates
four components of fields, thus, in principle we have four transformations. In order to reduce the number of degrees of freedom,
our starting point will be the Yang-Mills Lagrangian in the light-cone gauge, on the constant light-cone time $x^+$. As we shall recall below, such formulation reduces the four components of the gauge field $\hat{A}^{\mu}=t^aA_a^{\mu}$ to just two. Here, $t^a$ are  color generators in the fundamental representation.\footnote{We use the following normalization of the color generators, common in amplitude-related literature: $\left[t^{a},t^{b}\right]=i\sqrt{2}f^{abc}t^{c}$ and $\mathrm{Tr}(t^{a}t^{b}) = \delta^{ab}$. We re-scale the coupling constant as $g\rightarrow g/\sqrt{2}$ to accommodate for the additional factors of $\sqrt{2}$ resulting in that normalization.} To this end, we introduce the "plus" and "minus" light-cone coordinates for a four-vector $v^{\mu}$ 
\begin{equation}
v^{+}=v\cdot\eta\,, \,\,\,\, v^{-}=v\cdot\widetilde{\eta}\,,\label{eq:LCplusminus}
\end{equation}
where $\eta=\left(1,0,0,-1\right)/\sqrt{2}$, $\widetilde{\eta}=\left(1,0,0,1\right)/\sqrt{2}$, and two transverse coordinates
\begin{equation}
v^{\bullet}=v\cdot\varepsilon_{\bot}^{+}\,,\,\,\,\, v^{\star}=v\cdot\varepsilon_{\bot}^{-}\,,\label{eq:LCbulletstar}
\end{equation}
defined by the complex null vectors
$\varepsilon_{\perp}^{\pm}=\left(0,1,\pm i,0\right)/\sqrt{2}$. The scalar product of two four-vectors in these coordinates reads $u\cdot w=u^{+}w^{-}+u^{-}w^{+}-u^{\bullet}w^{\star}-u^{\star}w^{\bullet}$. In order to lower the indices one needs to flip $+\leftrightarrow -$ and $\star \leftrightarrow  \bullet$, where the latter operation also causes a sign change.

The light-cone Yang-Mills action is obtained by setting the light-cone gauge $\hat{A}^+=0$ and integrating out the $\hat{A}^-$ field from the partition function. The resulting action has only two transverse fields $\hat{A}^{\bullet}$ and $\hat{A}^{\star}$ and reads \cite{Scherk1975}
\begin{multline}
S_{\mathrm{Y-M}}^{\left(\mathrm{LC}\right)}\left[A^{\bullet},A^{\star}\right]=\int dx^{+}\int d^{3}\mathbf{x}\,\,\Bigg\{ 
-\mathrm{Tr}\,\hat{A}^{\bullet}\square\hat{A}^{\star}
-2ig\,\mathrm{Tr}\,\partial_{-}^{-1}\partial_{\bullet} \hat{A}^{\bullet}\left[\partial_{-}\hat{A}^{\star},\hat{A}^{\bullet}\right] \\
-2ig\,\mathrm{Tr}\,\partial_{-}^{-1}\partial_{\star}\hat{A}^{\star}\left[\partial_{-}\hat{A}^{\bullet},\hat{A}^{\star}\right]
-2g^{2}\,\mathrm{Tr}\,\left[\partial_{-}\hat{A}^{\bullet},\hat{A}^{\star}\right]\partial_{-}^{-2}\left[\partial_{-}\hat{A}^{\star},\hat{A}^{\bullet}\right]
\Bigg\}
\,,\label{eq:YM_LC_action}
\end{multline}
where the bold position-space three-vector is defined as $\mathbf{x}\equiv\left(x^{-},x^{\bullet},x^{\star}\right)$ so that $x=(x^+,\mathbf{x})$. The nabla operator reads  $\square=2(\partial_+\partial_- - \partial_{\bullet}\partial_{\star})$ in these coordinates.  The presence of just the physical degrees of freedom in the action leads to a natural identification of the helicity content of the vertices. Assigning the "plus" helicity to $\hat{A}^{\bullet}$ field , and the "minus" helicity to the field $\hat{A}^{\star}$, we see that there are $(++-)$, $(--+)$ and $(++--)$ vertices in the action.

Following the idea of \cite{Mansfield2006}, we look for a field transformation
\begin{equation}
    \left\{\hat{A}^{\bullet},\hat{A}^{\star}\right\} \rightarrow \Big\{\hat{Z}^{\bullet}\big[{A}^{\bullet},{A}^{\star}\big],\hat{Z}^{\star}\big[{A}^{\bullet},{A}^{\star}\big]\Big\} \, ,
    \label{eq:general_transf}
\end{equation}
which maps the kinetic term and both the triple-gluon vertices into a free term in the new action. In addition, we demand that the transformation is canonical, so that the functional measure in the partition function is preserved, up to a field independent factor. 

Before we present the details on the transformation facilitating the above requirements, let us make some introductory remarks. Suppose we have a set of generalized coordinates and momenta $q_i$, $p_i$. Consider a canonical transformation to a new set $Q_i$, $P_i$. 
Consider now a particular generating function $\mathcal{G}$ for the canonical transformation between $\{q,p\}$ and $\{Q,P\}$,  depending only on 
the
generalized coordinates, $\mathcal{G}(q,Q)$. Then, the relation between the original and the transformed coordinates is
\begin{equation}
     p_i =  \frac{\partial\mathcal{G}(q,Q)}{\partial q_i} \,, \qquad 
     P_i = - \frac{\partial\mathcal{G}(q,Q)}{\partial Q_i} \,.
    \label{eq:generatingfunc_1}
\end{equation}

In our context, the role of $q_i$ coordinate is played by the $\hat{A}^{\bullet}(x)$ field, and the canonical momentum $p_i$ is $\partial_{-}\hat{A}^{\star}(x)$. In the new theory, we identify $Q_i$ with $\hat{Z}^{\star}(x)$ and $P_i$ with $\partial_{-}\hat{Z}^{\bullet}(x)$. Therefore, the analogous relations are 
\begin{equation}
     \partial_{-}A^{\star}_a(x^+,\mathbf{y}) =  \frac{\delta \, \mathcal{G}[A^{\bullet},Z^{\star} ](x^+)}{\delta A_a^{\bullet}\left(x^+,\mathbf{y}\right)} \,, \qquad 
     \partial_{-}Z^{\bullet}_a(x^+,\mathbf{y}) = - \frac{\delta \, \mathcal{G}[A^{\bullet},Z^{\star} ](x^+)}{\delta Z_a^{\star}\left(x^+,\mathbf{y}\right)} \,,
     \label{eq:generatingfunc2}
\end{equation}
where we have explicitly denoted the fact, that the transformation is performed on the hyper-surface of constant light-cone time $x^+$.
Although the transformation \eqref{eq:general_transf} is rather complicated, we found that, quite amazingly, the generating functional $\mathcal{G}[A^{\bullet},Z^{\star}]$ can be written in the following simple form:
\begin{equation}
    \mathcal{G}[A^\bullet,Z^\star](x^+) =
    -\int\! d^3\mathbf{x}\,\,\,\Tr\,
     \hat{\mathcal{W}}^{\,-1}_{(-)}[Z](x)\,\,
     \partial_- \hat{\mathcal{W}}_{(+)}[A](x) \,,
    \label{eq:generatingfunc3}
\end{equation}
where the functional $\mathcal{W}_{(\pm)}[K]$, for a generic vector field $K^{\mu}$, is directly related to the straight infinite Wilson line in the following way:
\begin{equation}
     \mathcal{W}^{a}_{(\pm)}[K](x)=\int_{-\infty}^{\infty}d\alpha\,\mathrm{Tr}\left\{ \frac{1}{2\pi g}t^{a}\partial_{-}\, \mathbb{P}\exp\left[ig\int_{-\infty}^{\infty}\! ds\, \varepsilon_{\alpha}^{\pm}\cdot \hat{K}\left(x+s\varepsilon_{\alpha}^{\pm}\right)\right]\right\} \, ,
\label{eq:WL_gen}
\end{equation}
with 
\begin{equation}
    \varepsilon_{\alpha}^{\pm\, \mu} = \varepsilon_{\perp}^{\pm\, \mu }- \alpha \eta^{\mu} \, .
    \label{eq:epsilon_alpha}
\end{equation}
The above four vector  has the form of a gluon polarization vector. Indeed for $\alpha=p\cdot\varepsilon_{\perp}^{\pm}/p^{+}$, it is the transverse polarization vector for a gluon with momentum $p$. This type of functional has been used for the first time in \cite{Kotko2017} in the context of the MHV Lagrangian. 
The inverse functional to the Wilson line, $\mathcal{W}^{-1}$, is defined 
using the relation
$\mathcal{W}[\mathcal{W}^{-1}[K]]=K$. 
Note, that the functionals $\mathcal{W}_{(+)}$ and $\mathcal{W}_{(-)}$ are not exactly hermitian conjugates of each other; only the projection on $\varepsilon_{\alpha}^{+}$ or $\varepsilon_{\alpha}^{-}$ changes inside the path-ordered exponential, but the sign of $ig$ remains unchanged.

In the following sections we shall present more details on the  implication of the transformation given by \eqref{eq:generatingfunc2} and the exact form of the vertices in the new action. In the remaining part of this section, we will outline the general structure of the new action.

From Eqs.~\eqref{eq:generatingfunc2}-\eqref{eq:generatingfunc3} one can see that the fields $\hat{A}^{\bullet}$ and $\hat{A}^{\star}$ have the following 
general
expansion in terms of the new fields:
\begin{equation}
    A_a^{\bullet}(x^+;\mathbf{x})=\sum_{n=1}^{\infty}
    \int\! d^3\mathbf{y}_1\dots d^3\mathbf{y}_n \sum_{i=1}^{n}\, \Xi_{i,n-i}^{ab_1\dots b_n}(\mathbf{x};\mathbf{y}_1,\dots,\mathbf{y}_n) \prod_{k=1}^{i}Z_{b_k}^{\bullet}(x^+;\mathbf{y}_k)
    \prod_{l=i+1}^{n}Z_{b_l}^{\star}(x^+;\mathbf{y}_l) \,,
    \label{eq:Abullet_to_Z}
\end{equation}
\begin{equation}
    A_a^{\star}(x^+;\mathbf{x})=\sum_{n=1}^{\infty}
    \int\! d^3\mathbf{y}_1\dots d^3\mathbf{y}_n \sum_{i=1}^{n}\, \Lambda_{i,n-i}^{ab_1\dots b_n}(\mathbf{x};\mathbf{y}_1,\dots,\mathbf{y}_n) \prod_{k=1}^{i}Z_{b_k}^{\star}(x^+;\mathbf{y}_k)
    \prod_{l=i+1}^{n}Z_{b_l}^{\bullet}(x^+;\mathbf{y}_l) \,,
    \label{eq:Astar_to_Z}
\end{equation}
where $\Xi_{i,j}^{ab_1\dots b_{i+j}}(\mathbf{x};\mathbf{y}_1\dots \mathbf{y}_{i+j})$  is  an apriori unknown kernel for  $i$ number of  $Z^{\bullet}$ fields and $j$ number of $Z^{\star}$  fields in the expansion of $\hat{A}^{\bullet}$, depending on the adjoint color indices $a,b_1\dots b_{i+j}$ and not depending on the light-cone time. Similarly, $\Lambda_{i,j}^{ab_1\dots b_{i+j}}(\mathbf{x};\mathbf{y}_1\dots \mathbf{y}_{i+j})$ is the kernel for $i$ number of  $Z^{\star}$ fields and $j$ number of $Z^{\bullet}$  fields in the expansion of $\hat{A}^{\star}$. At lowest order we must have
\begin{equation}
    A_a^{\bullet}(x^+;\mathbf{x})= Z_{a}^{\bullet}(x^+;\mathbf{x})+\dots \,\,, \qquad
    A_a^{\star}(x^+;\mathbf{x})=
    Z_{a}^{\star}(x^+;\mathbf{x})+\dots \,\,.
    \label{eq:A_to_Z_zeroth}
\end{equation}

In principle, one could find explicitly the kernels $\Xi_{i,j}$, $\Lambda_{i,j}$ from Eqs.~\eqref{eq:generatingfunc2}-\eqref{eq:generatingfunc3}. However, as we demonstrate in the next section, there is a much better way of doing that, which utilizes the existing results on the MHV Lagrangian \cite{Kotko2017,Kakkad2020}. Since we want to describe a general structure of the action, for the rest of this section we shall assume that the kernels are known.

Inserting the solutions \eqref{eq:Abullet_to_Z}-\eqref{eq:Astar_to_Z} to the Yang-Mills action \eqref{eq:YM_LC_action}, we find the following structure of the new action: 
\begin{align}
S_{\mathrm{Y-M}}^{\left(\mathrm{LC}\right)}\left[Z^{\bullet},Z^{\star}\right] =\int dx^{+} \Bigg\{ & 
-\int d^{3}\mathbf{x}\,\mathrm{Tr}\,\hat{Z}^{\bullet}\square\hat{Z}^{\star} \nonumber \\
 & + \mathcal{L}^{(\mathrm{LC})}_{--++}+ \mathcal{L}^{(\mathrm{LC})}_{--+++}+\mathcal{L}^{(\mathrm{LC})}_{--++++} + \dots \nonumber \\
& + \mathcal{L}^{(\mathrm{LC})}_{---++}+ \mathcal{L}^{(\mathrm{LC})}_{---+++}+\mathcal{L}^{(\mathrm{LC})}_{---++++} + \dots \nonumber \\
%& + \mathcal{L}^{(\mathrm{LC})}_{----++}+ \mathcal{L}^{(\mathrm{LC})}_{----+++}+\mathcal{L}^{(\mathrm{LC})}_{----++++}+ \dots \\
%& + \dots \\
& \,\, \vdots \nonumber \\
& + \mathcal{L}^{(\mathrm{LC})}_{---\dots -++}+ \mathcal{L}^{(\mathrm{LC})}_{---\dots -+++}+\mathcal{L}^{(\mathrm{LC})}_{---\dots -++++}+ \dots
\Bigg\}
\,,\label{eq:Z_action1}
\end{align}
where the $n$-point interaction vertex, $n\geq 4$, that couples $m$ minus helicity fields, $m\geq 2$, and $n-m$ plus helicity fields, has the following general form: 
\begin{equation}
    \mathcal{L}_{\underbrace{-\,\cdots\,-}_{m}\underbrace{+ \,\cdots\, +}_{n-m}}^{\left(\mathrm{LC}\right)}= 
   \int\!d^{3}\mathbf{y}_{1}\dots d^{3}\mathbf{y}_{n} \,\, \mathcal{U}^{b_1 \dots b_{n}}_{-\dots-+\dots+}\left(\mathbf{y}_{1},\cdots \mathbf{y}_{n}\right) 
   \prod_{i=1}^{m}Z^{\star}_{b_i} (x^+;\mathbf{y}_{i})
   \prod_{j=1}^{n-m}Z^{\bullet}_{b_j} (x^+;\mathbf{y}_{j}) \, .
   \label{eq:Z_vertex_lagr_pos}
\end{equation}

The above action has the following properties, which we will elaborate on in the next sections:
\begin{enumerate}[label={\it\roman*}$\,$)]%,leftmargin=35pt]
    \item There are no three point interaction vertices. 
    \item At the classical level there are no all-plus, all-minus, as well as $(-+\dots +)$, $(- \dots - +)$ vertices.
    \item There are MHV vertices, $(--+\dots +)$, corresponding to MHV amplitudes in the on-shell limit.
    \item There are $\overline{\mathrm{MHV}}$ vertices, $(-\dots - ++)$, corresponding to $\overline{\mathrm{MHV}}$ amplitudes in the on-shell limit.
    \item All vertices have the form which can be easily calculated.
\end{enumerate}
 Because the lowest vertex is the single MHV four-point vertex that corresponds to the four-gluon MHV amplitude in the on-shell limit, the new action provides an efficient way to construct tree amplitudes with high multiplicity of legs, as we will demonstrate later in Section~\ref{sec:Applications} by computing several examples.

%-----------------------------------------------
\subsection{Derivation}

As we shall see in the following, the easiest way to derive the action \eqref{eq:Z_action1} from the Yang-Mills action \eqref{eq:YM_LC_action} is to first transform the latter into an action containing the MHV vertices. Thus, we start by a brief summary of this procedure.

%...............................................
\subsubsection{MHV action}

As explained in detail in \cite{Mansfield2006}, the MHV action implementing the CSW rules \cite{Cachazo2004} is obtained by performing a canonical field transformation with a requirement that the kinetic term and the $(++-)$ triple-gluon vertex is mapped to a free kinetic term in the new action:
\begin{equation}
\mathrm{Tr}\,\hat{A}^{\bullet}\square\hat{A}^{\star}
+2ig\,\mathrm{Tr}\,\partial_{-}^{-1}\partial_{\bullet} \hat{A}^{\bullet}\left[\partial_{-}\hat{A}^{\star},\hat{A}^{\bullet}\right]
\,\, \longrightarrow \,\,
\mathrm{Tr}\,\hat{B}^{\bullet}\square\hat{B}^{\star}
\,.\label{eq:MansfieldTransf1}
\end{equation}
Note that, the two terms on the l.h.s constitute the self-dual sector of the Yang-Mills theory \cite{Bardeen1996,Chalmers1996,Cangemi1997,Rosly1997,Monteiro2011}. Therefore, as shown in \cite{Kakkad2020} the solution to the required transformation of fields can be expressed in terms of the straight infinite Wilson line lying on the self-dual plane, i.e. the plane spanned by the $\varepsilon_{\perp}^+$ and $\eta$. It is exactly the Wilson line $\mathcal{W}_{(+)}$ introduced in the   preceding section.  The new fields $B$, expressed in terms of the Yang-Mills fields $A$, read
\begin{equation}
    B^{\bullet}_a[A^{\bullet}](x)=\mathcal{W}_{(+)}^a[A](x)\,,\qquad
    B_a^{\star}[A^{\bullet},A^{\star}](x) = 
    \int\! d^3\mathbf{y} \,
     \left[ \frac{\partial^2_-(y)}{\partial^2_-(x)} \,
     \frac{\delta \mathcal{W}^a_{(+)}[A](x^+;\mathbf{x})}{\delta {A}_c^{\bullet}(x^+;\mathbf{y})} \right] 
     {A}_c^{\star}(x^+;\mathbf{y})
      \, .
      \label{eq:Bfield_transform}
\end{equation}
The expressions for the fields in the momentum space have the following form \cite{Kotko2017,Kakkad2020}
\begin{equation}
    \widetilde{B}^{\bullet}_a(x^+;\mathbf{P}) = \sum_{n=1}^{\infty} 
    \int d^3\mathbf{p}_1\dots d^3\mathbf{p}_n \, \widetilde{\Gamma}_n^{a\{b_1\dots b_n\}}(\mathbf{P};\{\mathbf{p}_1,\dots ,\mathbf{p}_n\}) \prod_{i=1}^n\widetilde{A}^{\bullet}_{b_i}(x^+;\mathbf{p}_i)\,,
    \label{eq:B_bull_exp}
\end{equation}
\begin{equation}
    \widetilde{B}^{\star}_a(x^+;\mathbf{P}) = \sum_{n=1}^{\infty} 
    \int d^3\mathbf{p}_1\dots d^3\mathbf{p}_n \, {\widetilde \Upsilon}_{n}^{a b_1 \left \{b_2 \dots b_n \right \} }(\mathbf{P}; \mathbf{p_1} ,\left \{ \mathbf{p_2} , \dots ,\mathbf{p_n} \right \}) \widetilde{A}^{\star}_{b_1}(x^+;\mathbf{p}_1)\prod_{i=2}^n\widetilde{A}^{\bullet}_{b_i}(x^+;\mathbf{p}_i)\,,
    \label{eq:B_star_exp}
\end{equation}
where
\begin{equation}
    \widetilde{\Gamma}^{a\{b_1\dots b_n\}}_n (\mathbf{P};\{\mathbf{p}_1,\dots,\mathbf{p}_n\}) = (-g)^{n-1} \frac{\delta^3\left(\mathbf{p}_1+\dots+\mathbf{p}_n-\mathbf{P}\right)\,\Tr\!\left(t^at^{b_1}\dots t^{b_n}\right)}{\widetilde{v}^*_{1(1\cdots n)} \widetilde{v}^*_{(12)(1\cdots n)} \cdots \widetilde{v}^*_{(1 \cdots n-1)(1\cdots n)}} \, ,
    \label{eq:Gamma_n}
\end{equation}
\begin{equation}
    {\widetilde \Upsilon}_{n}^{a b_1 \left \{b_2 \cdots b_n \right \} }(\mathbf{P}; \mathbf{p_1} ,\left \{ \mathbf{p_2} , \dots ,\mathbf{p_n} \right \}) = n\left(\frac{p_1^+}{p_{1\cdots n}^+}\right )^2 {\widetilde \Gamma}_{n}^{a b_1 \dots b_n }(\mathbf{P}; \mathbf{p_{1}},  \dots ,\mathbf{p_{n}} ) \, .
    \label{eq:Upsilon_n}
\end{equation}
Above, the tildes over the fields and the kernels $\Gamma$, $\Upsilon$ denote the Fourier transformed quantities with respect to the three momenta $\mathbf{p}=(p^+,p^{\bullet},p^{\star})$. The curly brackets denote the symmetrization with respect to the pairs of momentum and color indices. The $\widetilde{v}_{ij}$, $\widetilde{v}^{\star}_{ij}$ are quantities similar to spinor products $\left<ij\right>$, $\left[ij\right]$, with the following explicit definitions (first introduced in \cite{Motyka2009} in the context of the gluon wave function):
\begin{equation}
    \widetilde{v}_{ij}=
    p_i^+\left(\frac{p_{j}^{\star}}{p_{j}^{+}}-\frac{p_{i}^{\star}}{p_{i}^{+}}\right), \qquad 
\widetilde{v}^*_{ij}=
    p_i^+\left(\frac{p_{j}^{\bullet}}{p_{j}^{+}}-\frac{p_{i}^{\bullet}}{p_{i}^{+}}\right)\, .
\label{eq:vtilde}
\end{equation}
They appear quite naturally in the Wilson line approach, because 
\begin{equation}
    \widetilde{v}^*_{ij}=-(\varepsilon_i^+\cdot p_j)\,, \quad 
    \widetilde{v}_{ij}=-(\varepsilon_i^-\cdot p_j )\, ,
    \label{eq:vtilde1}
\end{equation}
where $\varepsilon^{\pm}_i$ is the polarization vector for a momentum $p_i$ obtained from \eqref{eq:epsilon_alpha} which  appears as the direction of the Wilson line. See \cite{Cruz-Santiago2015} for several useful properties of the $\widetilde{v}_{ij}$ symbols.
We also use a shorthand notation for the sum of momenta $p_1+\dots +p_n\equiv p_{1\dots n}$.

The expressions with Wilson lines \eqref{eq:Bfield_transform} (or equivalently \eqref{eq:B_bull_exp}, \eqref{eq:B_star_exp}) can be inverted to obtain the power expansions for $A^{\bullet}$, $A^{\star}$ in terms of $B^{\bullet}$, $B^{\star}$, which are consistent with \cite{Ettle2006b}. In the momentum space we get:
\begin{equation}
    \widetilde{A}^{\bullet}_a(x^+;\mathbf{P}) = \sum_{n=1}^{\infty} 
    \int d^3\mathbf{p}_1\dots d^3\mathbf{p}_n \, \widetilde{\Psi}_n^{a\{b_1\dots b_n\}}(\mathbf{P};\{\mathbf{p}_1,\dots ,\mathbf{p}_n\}) \prod_{i=1}^n\widetilde{B}^{\bullet}_{b_i}(x^+;\mathbf{p}_i)\,,
    \label{eq:A_bull_exp}
\end{equation}
\begin{equation}
    \widetilde{A}^{\star}_a(x^+;\mathbf{P}) = \sum_{n=1}^{\infty} 
    \int d^3\mathbf{p}_1\dots d^3\mathbf{p}_n \, {\widetilde \Omega}_{n}^{a b_1 \left \{b_2 \cdots b_n \right \} }(\mathbf{P}; \mathbf{p_1} ,\left \{ \mathbf{p_2} , \dots ,\mathbf{p_n} \right \}) \widetilde{B}^{\star}_{b_1}(x^+;\mathbf{p}_1)\prod_{i=2}^n\widetilde{B}^{\bullet}_{b_i}(x^+;\mathbf{p}_i)\, ,
    \label{eq:A_star_exp}
\end{equation}
where the kernels are
\begin{equation}
    {\widetilde \Psi}_{n}^{a \left \{b_1 \cdots b_n \right \} }(\mathbf{P}; \left \{\mathbf{p}_{1},  \dots ,\mathbf{p}_{n} \right \}) =- (-g)^{n-1} \,\,  
    \frac{{\widetilde v}^{\star}_{(1 \cdots n)1}}{{\widetilde v}^{\star}_{1(1 \cdots n)}} \, 
    \frac{\delta^{3} (\mathbf{p}_{1} + \cdots +\mathbf{p}_{n} - \mathbf{P})\,\,  \mathrm{Tr} (t^{a} t^{b_{1}} \cdots t^{b_{n}})}{{\widetilde v}^{\star}_{21}{\widetilde v}^{\star}_{32} \cdots {\widetilde v}^{\star}_{n(n-1)}}  
      \, ,
    \label{eq:psi_kernel}
\end{equation}
\begin{equation}
    {\widetilde \Omega}_{n}^{a b_1 \left \{b_2 \cdots b_n \right \} }(\mathbf{P}; \mathbf{p}_{1} , \left \{ \mathbf{p}_{2} , \dots ,\mathbf{p}_{n} \right \} ) = n \left(\frac{p_1^+}{p_{1\cdots n}^+}\right)^2 {\widetilde \Psi}_{n}^{a b_1 \cdots b_n }(\mathbf{P};  \mathbf{p}_{1},  \dots , \mathbf{p}_{n}) \, .
    \label{eq:omega_star_z}
\end{equation}
Inserting the above solutions to the Yang-Mills action we can derive the MHV action:
\begin{equation}
S_{\mathrm{Y-M}}^{\left(\mathrm{LC}\right)}\left[{B}^{\bullet}, {B}^{\star}\right]=\int dx^{+}\left(
-\int d^{3}\mathbf{x}\,\mathrm{Tr}\,\hat{B}^{\bullet}\square\hat{B}^{\star} 
+\mathcal{L}_{--+}^{\left(\mathrm{LC}\right)}+\dots+\mathcal{L}_{--+\dots+}^{\left(\mathrm{LC}\right)}+\dots\right)\,,\label{eq:MHV_action}
\end{equation}
where the $n$-point MHV interaction terms are
\begin{multline}
\mathcal{L}_{--+\dots+}^{\left(\mathrm{LC}\right)}=\int d^{3}\mathbf{p}_{1}\dots d^{3}\mathbf{p}_{n}\delta^{3}\left(\mathbf{p}_{1}+\dots+\mathbf{p}_{n}\right)\,
\widetilde{\mathcal{V}}_{--+\dots+}^{b_{1}\dots b_{n}}\left(\mathbf{p}_{1},\dots,\mathbf{p}_{n}\right)
\\ \widetilde{B}_{b_{1}}^{\star}\left(x^+;\mathbf{p}_{1}\right)\widetilde{B}_{b_{2}}^{\star}\left(x^+;\mathbf{p}_{2}\right)\widetilde{B}_{b_{3}}^{\bullet}\left(x^+;\mathbf{p}_{3}\right)\dots\widetilde{B}_{b_{n}}^{\bullet}\left(x^+;\mathbf{p}_{n}\right)
\,,
\label{eq:MHV_n_point}
\end{multline}
with the MHV vertices 
\begin{equation}
\widetilde{\mathcal{V}}_{--+\dots+}^{b_{1}\dots b_{n}}\left(\mathbf{p}_{1},\dots,\mathbf{p}_{n}\right)= \!\!\sum_{\underset{\text{\scriptsize permutations}}{\text{noncyclic}}}
 \mathrm{Tr}\left(t^{b_1}\dots t^{b_n}\right)
 \mathcal{V}\left(1^-,2^-,3^+,\dots,n^+\right)
\,,
\label{eq:MHV_vertex}
\end{equation}
where the color ordered vertex reads 
\begin{equation}
\mathcal{V}\left(1^-,2^-,3^+,\dots,n^+\right)= 
\frac{(-g)^{n-2}}{(n-2)!}  \left(\frac{p_{1}^{+}}{p_{2}^{+}}\right)^{2}
\frac{\widetilde{v}_{21}^{*4}}{\widetilde{v}_{1n}^{*}\widetilde{v}_{n\left(n-1\right)}^{*}\widetilde{v}_{\left(n-1\right)\left(n-2\right)}^{*}\dots\widetilde{v}_{21}^{*}}
\,.
\label{eq:MHV_vertex_colororder}
\end{equation}

Above, we defined the momentum space color ordered vertex without the tilde, as this will not lead to any confusion.
Note that, we have written the MHV vertices in a form, where the negative helicity fields are always adjacent, but there is a sum over the color permutations, together with the proper symmetry factor.

The vertices \eqref{eq:MHV_vertex_colororder} are fully off-shell quantities, but they correspond to the MHV amplitudes in the on-shell limit (which is evident from the fact that the $\widetilde{v}_{ij}$ symbols are in one-to-one correspondence to the spinor products). These vertices are in general not gauge invariant for off-shell kinematics, but as shown in \cite{Cruz-Santiago2015,Kotko2017}, they do constitute a gauge invariant off-shell amplitude, when only one plus-helicity leg is kept off-shell.

%...............................................
\subsubsection{Canonical transformation of the MHV action}

In \cite{Kakkad2020} we argued, that the second equation in  \eqref{eq:Bfield_transform} suggests, that there should exist a more general structure 
spanning not only on the 
 $\varepsilon^+_{\perp}\text{-}\eta$ plane. The latter should be just a slice of this more general structure. That is, there should exist a functional which path-orders the $A^{\star}$ fields in the plane perpendicular to the $\varepsilon_{\perp}^+\text{-}\eta$ plane.

In order to actually introduce such an object, let us consider a canonical transformation of the MHV action itself. We demand that
\begin{equation}
\mathcal{L}_{-+}[B^{\bullet},B^{\star}]+\mathcal{L}_{--+}[B^{\bullet},B^{\star}]
\,\, \longrightarrow \,\,
\mathcal{L}_{-+}[Z^{\bullet},Z^{\star}]
\,,
\label{eq:BtoZtransform}
\end{equation}
where $\mathcal{L}_{-+}$ is just the kinetic term in either $B$ or $Z$ fields, cf. Eq.~\eqref{eq:MansfieldTransf1}.
The vertex $\mathcal{L}_{--+}[B^{\bullet},B^{\star}]$ appearing in \eqref{eq:MHV_action} has exactly the same form as the $(--+)$ triple-gluon vertex in the original Yang-Mills action \eqref{eq:YM_LC_action}, but with $A$ fields replaced by $B$ fields. Therefore, the corresponding transformations are analogous to those leading to the MHV action, but with the replacement $\bullet \leftrightarrow \star$. More precisely
\begin{equation}
    Z^{\star}_a[B^{\star}](x)=\mathcal{W}_{(-)}^a[B](x)\,,\qquad
    Z_a^{\bullet}[B^{\bullet},B^{\star}](x) = 
    \int\! d^3\mathbf{y} \,
     \left[ \frac{\partial^2_-(y)}{\partial^2_-(x)} \,
     \frac{\delta \mathcal{W}^a_{(-)}[B](x^+;\mathbf{x})}{\delta {B}_c^{\star}(x^+;\mathbf{y})} \right] 
     {B}_c^{\bullet}(x^+;\mathbf{y})
      \, .
      \label{eq:Zfield_transform}
\end{equation}

Let us point out an important feature of the above formula. Unlike the transformation leading to the MHV action, Eq.~\eqref{eq:Bfield_transform}, which involved the Wilson lines $\mathcal{W}_{(+)}$ along $\varepsilon_{\alpha}^+$, here we have the Wilson line $\mathcal{W}_{(-)}$ that have directions $\varepsilon_{\alpha}^-$, see the definitions \eqref{eq:WL_gen}. Thus, pictorially, the $Z^{\star}$ field is the Wilson line on the $\eta\text{-}\varepsilon_{\perp}^{-}$ plane, where the path ordered fields are themselves Wilson lines on the $\eta\text{-}\varepsilon_{\perp}^{+}$ plane (see also Fig.~\ref{fig:geometry} in Section~\ref{sec:Summary}).

Already at this stage one can check that the generating functional \eqref{eq:generatingfunc3} is consistent with the above transformations. 
Inserting \eqref{eq:generatingfunc3} to \eqref{eq:generatingfunc2} we have
\begin{equation}
\partial_{-}A_{a}^{\star}\left(x^+;\mathbf{x}\right)=-\int d^{3}\mathbf{y}\,\, \mathcal{W}^{c\,\,-1}_{(-)}[Z](x^+;\mathbf{y})\, %\left[
\partial_{-}\,\frac{\delta}{\delta A_{a}^{\bullet}\left(x^+;\mathbf{x}\right)} \mathcal{W}^{c}_{(+)}[A](x^+;\mathbf{y})
%\right]
\,,
\label{eq:Transformation_A-_Z}
\end{equation}
and 
\begin{equation}
\partial_{-}Z_{a}^{\bullet}\left(x^+;\mathbf{x}\right)=\int d^{3}\mathbf{y}\,\left[\frac{\delta }{\delta Z_{a}^{\star}\left(x^+;\mathbf{x}\right)}
\mathcal{W}^{c\,\,-1}_{(-)}[Z](x^+;\mathbf{y}) \right]
\,\, \partial_- \mathcal{W}^{c}_{(+)}[A](x^+;\mathbf{y}) \, .
\label{eq:Transformation_Z+}
\end{equation}
Integrating Eq.~\eqref{eq:Transformation_A-_Z} by parts and using the first equation of \eqref{eq:Bfield_transform} we get
\begin{equation}
    \partial_{-}A_{a}^{\star}\left(x^+;\mathbf{x}\right)=\int d^{3}\mathbf{y}\,
    \left[\partial_{-}\mathcal{W}^{c\,\,-1}_{(-)}[Z](x^+;\mathbf{y})\right]\frac{\delta B_{c}^{\bullet}\left(x^+;\mathbf{y}\right)}{\delta A_{a}^{\bullet}\left(x^+;\mathbf{x}\right)}
    \, .
    \label{eq:Z_wl_src}
\end{equation}
Comparing this with the canonical transformation rule for the $A^{\star}$ field of \cite{Mansfield2006} which reads
\begin{equation}
\partial_{-}A_{a}^{\star}\left(x^+;\mathbf{x}\right)=\int d^{3}\mathbf{y}\,\frac{\delta B_{c}^{\bullet}\left(x^+;\mathbf{y}\right)}{\delta A_{a}^{\bullet}\left(x^+;\mathbf{x}\right)}
\, \partial_{-}B_{c}^{\star}\left(x^+;\mathbf{y}\right) \, ,
\label{eq:Transformation_A-}
\end{equation}
we see that 
\begin{equation}
   B_{c}^{\star}[Z^{\star}](x) = \mathcal{W}^{c\,\,-1}_{(-)}[Z](x) \,, 
   \label{eq:B_star_Z}
\end{equation}
 or, upon inverting, 
\begin{equation}
   Z_{c}^{\star}[B^{\star}](x) = \mathcal{W}^{c}_{(-)}[B](x) \,, 
   \label{eq:B_star_Z_1}
\end{equation}
which gives the left equation of \eqref{eq:Zfield_transform}.
Inserting now \eqref{eq:B_star_Z} into \eqref{eq:Transformation_Z+} and using the first equation of  \eqref{eq:Bfield_transform} we get
\begin{equation}
\partial_{-}Z_{a}^{\bullet}\left(x^+;\mathbf{x}\right)=\int d^{3}\mathbf{y}\,\frac{\delta B_{c}^{\star}\left(x^+;\mathbf{y}\right)}{\delta Z_{a}^{\star}\left(x^+;\mathbf{x}\right)}
\, \partial_{-}B_{c}^{\bullet}\left(x^+;\mathbf{y}\right) \, .
\label{eq:Transformation_Z+_1}
\end{equation}
 or 
\begin{equation}
\partial_{-}B_{a}^{\bullet}\left(x^+;\mathbf{x}\right)=\int d^{3}\mathbf{y}\,\frac{\delta Z_{c}^{\star}\left(x^+;\mathbf{y}\right)}{\delta B_{a}^{\star}\left(x^+;\mathbf{x}\right)}
\, \partial_{-}Z_{c}^{\bullet}\left(x^+;\mathbf{y}\right) \, .
\label{eq:Transformation_Z+_2}
\end{equation}
which, is virtually the same as \eqref{eq:Transformation_A-}, but with the replacement $\bullet\leftrightarrow\star$ and the $Z$ fields instead of $B$ fields and $B$ fields instead of $A$ fields.
Since \eqref{eq:Transformation_A-}, together with the left equation of \eqref{eq:Bfield_transform}, leads to the solution given by the right equation of \eqref{eq:Bfield_transform}, we can argue that  \eqref{eq:Transformation_Z+_2}, together with \eqref{eq:B_star_Z_1}, leads to the right equation of \eqref{eq:Zfield_transform}.
Note that because of the replacement $\bullet\leftrightarrow\star$, the Wilson line $\mathcal{W}_{(+)}$ in \eqref{eq:Bfield_transform} becomes  $\mathcal{W}_{(-)}$ in \eqref{eq:Zfield_transform}.

\begin{figure}[h!]
    \centering
    \includegraphics[width=8cm]{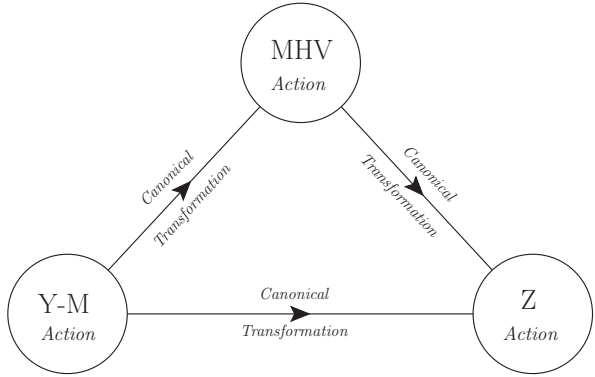}
    \caption{ 
    \small Two ways to derive the new action. First is the direct method which involves the generating functional \eqref{eq:generatingfunc3}. Second involves two consecutive canonical field transformation.
    } 
    \label{fig:CT_paths}
\end{figure}

To conclude, we have shown that the generating functional \eqref{eq:generatingfunc3} takes care of the chain of both canonical transformations, from $A$ fields to $B$ fields and from $B$ fields to $Z$ fields, simultaneously, as shown in the diagram in Fig.~\ref{fig:CT_paths}.

%...............................................
\subsubsection{Solution to the transformations}

We have just seen that the transformation from the Yang-Mills action to the new action generated by the functional \eqref{eq:generatingfunc3} is equivalent to two canonical transformations: first transforming the self-dual part of the Yang-Mills action to a free action in $B$-field theory, and then transforming the anti-self-dual part in the latter to a free term in the new $Z$-field theory. 
Therefore we can readily write the relations between the $Z$ fields and $B$ fields in momentum space. 

In order to derive the content of the $Z$-field action, we need to insert the expansions of $B$ fields in $Z$ fields.  
For the $B^{\star}$ field we find 
\begin{equation}
    \widetilde{B}^{\star}_a(x^+;\mathbf{P}) = \sum_{n=1}^{\infty} 
    \int d^3\mathbf{p}_1\dots d^3\mathbf{p}_n \, \overline{\widetilde{\Psi}}\,^{a\{b_1\dots b_n\}}_n(\mathbf{P};\{\mathbf{p}_1,\dots ,\mathbf{p}_n\}) \prod_{i=1}^n\widetilde{Z}^{\star}_{b_i}(x^+;\mathbf{p}_i)\,,
    \label{eq:BstarZ_exp}
\end{equation}
with
\begin{equation}
    \overline{\widetilde \Psi}\,^{a \left \{b_1 \cdots b_n \right \}}_{n}(\mathbf{P}; \left \{\mathbf{p}_{1},  \dots ,\mathbf{p}_{n} \right \}) =- (-g)^{n-1} \,\,  
    \frac{{\widetilde v}_{(1 \cdots n)1}}{{\widetilde v}_{1(1 \cdots n)}} \, 
    \frac{\delta^{3} (\mathbf{p}_{1} + \cdots +\mathbf{p}_{n} - \mathbf{P})\,\,  \mathrm{Tr} (t^{a} t^{b_{1}} \cdots t^{b_{n}})}{{\widetilde v}_{21}{\widetilde v}_{32} \cdots {\widetilde v}_{n(n-1)}}  
      \, .
    \label{eq:psiBar_kernel}
\end{equation}
Note that, the above quantity has the same form as \eqref{eq:psi_kernel}, however with $\widetilde{v}_{ij}^{\star}$ replaced by its complex conjugate $\widetilde{v}_{ij}$.
The expansion for the $B^{\bullet}$ field follows from \eqref{eq:A_star_exp} and reads
\begin{equation}
    \widetilde{B}^{\bullet}_a(x^+;\mathbf{P}) = \sum_{n=1}^{\infty} 
    \int d^3\mathbf{p}_1\dots d^3\mathbf{p}_n \, \overline{\widetilde \Omega}\,^{a b_1 \left \{b_2 \cdots b_n \right \}}_{n}(\mathbf{P}; \mathbf{p_1} ,\left \{ \mathbf{p_2} , \dots ,\mathbf{p_n} \right \}) \widetilde{Z}^{\bullet}_{b_1}(x^+;\mathbf{p}_1)\prod_{i=2}^n\widetilde{Z}^{\star}_{b_i}(x^+;\mathbf{p}_i)\,,
    \label{eq:BbulletZ_exp}
\end{equation}
where
\begin{equation}
    \overline{\widetilde \Omega}\,^{a b_1 \left \{b_2 \cdots b_n \right \}}_{n}(\mathbf{P}; \mathbf{p}_{1} , \left \{ \mathbf{p}_{2} , \dots ,\mathbf{p}_{n} \right \} ) = n \left(\frac{p_1^+}{p_{1\cdots n}^+}\right)^2 \overline{\widetilde \Psi}\,^{a b_1 \cdots b_n }_{n}(\mathbf{P}; \mathbf{p}_{1},  \dots ,\mathbf{p}_{n}) \, .
    \label{eq:omegaBar_kernel}
\end{equation}

Inserting the above expansions to \eqref{eq:B_bull_exp}-\eqref{eq:B_star_exp} makes it in principle possible to derive an explicit form of the expansions between $A$ fields and $Z$ fields. One can thus explicitly find the kernels $\Xi_{i,j}$, $\Lambda_{i,j}$ introduced in Eqs.~\eqref{eq:Abullet_to_Z}-\eqref{eq:Astar_to_Z}. However, it turns out that this is not necessary. It is much more efficient to  insert the above expressions into the MHV action, as we shall do in the following.

%...............................................
\subsubsection{General form of the vertex}
\label{z_action_vertex}

Although, formally, both field transformations $A\rightarrow B$ and $B\rightarrow Z$ are such that they remove the triple-gluon vertices,
 it is interesting to see how they actually cancel in the Yang-Mills action. 
Therefore, in the Appendix~\ref{sec:AppA} we directly check that the first terms of the expansions \eqref{eq:Abullet_to_Z}-\eqref{eq:Astar_to_Z} indeed cancel both of the triple-gluon vertices in the Yang-Mills action.

The remaining terms are obtained by inserting the expansions \eqref{eq:BstarZ_exp},\eqref{eq:BbulletZ_exp} into the MHV vertices \eqref{eq:MHV_vertex}, for $n\geq 4$. 
We shall find a general expression for the vertex when the negative helicity fields are adjacent.
Without loosing the generality we shall focus on the color ordered vertex, defined as
\begin{equation}
    \mathcal{U}_{-\dots-+\dots+}^{b_{1}\dots b_{n}}\left(\mathbf{p}_{1},\dots,\mathbf{p}_{n}\right)= \!\!\sum_{\underset{\text{\scriptsize permutations}}{\text{noncyclic}}}
 \mathrm{Tr}\left(t^{b_1}\dots t^{b_n}\right)
 \mathcal{U}\left(1^-,\dots,m^-,(m+1)^+,\dots,n^+\right)
\,,
\label{eq:Zvertex_color_decomp}
\end{equation}
where we assumed there are $m$ minus helicity legs. 
We shall also need the color ordered versions of the kernels in the expansions \eqref{eq:BstarZ_exp}-\eqref{eq:BbulletZ_exp}. We define 
\begin{equation}
    \overline{\widetilde{\Psi}}\,^{a\{b_{1}\dots b_{m}\}}_m\left(\mathbf{P};\{\mathbf{p}_{1},\dots,\mathbf{p}_{m}\}\right)= \!\!\sum_{\underset{\text{\scriptsize permutations}}{\text{noncyclic}}}
 \mathrm{Tr}\left(t^{b_1}\dots t^{b_m}\right)
 \overline{\Psi}\left(1^-,\dots,m^-\right)
\,,
\label{eq:PsiBar_color_decomp}
\end{equation}
and
\begin{equation}
    \overline{\widetilde{\Omega}}\,^{ab_{1}\{b_2\dots b_{m}\}}_m\left(\mathbf{P};\mathbf{p}_{1},\{\mathbf{p}_{2},\dots,\mathbf{p}_{m}\}\right)= \!\!\sum_{\underset{\text{\scriptsize permutations}}{\text{noncyclic}}}
 \mathrm{Tr}\left(t^{b_1}\dots t^{b_m}\right)
 \overline{\Omega}\left(1^+,2^-,\dots,m^-\right)
\,,
\label{eq:OmegaBar_color_decomp}
\end{equation}
where for further convenience we explicitly denoted the helicity of the legs in the color ordered kernels.
Note, 
 that the $\overline{\Psi}_m$ kernel multiplicates the minus helicity leg into $m$ minus helicity legs, whereas the $\overline{\Omega}_m$ kernel multiplicates the plus helicity leg into one plus helicity leg and $(m-1)$ adjacent  minus helicity legs.
Note also, that similar to \eqref{eq:MHV_vertex_colororder}, we have omitted the tilde signs in the momentum space color-ordered vertex and kernels.

\begin{figure}
    \centering
    \parbox[c]{4cm}{ \includegraphics[width=4cm]{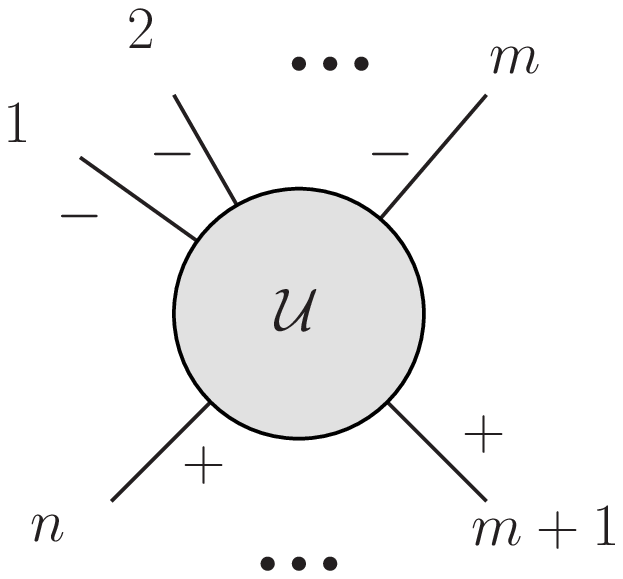}}\qquad
    \parbox[c]{8cm}{
    \includegraphics[width=8cm]{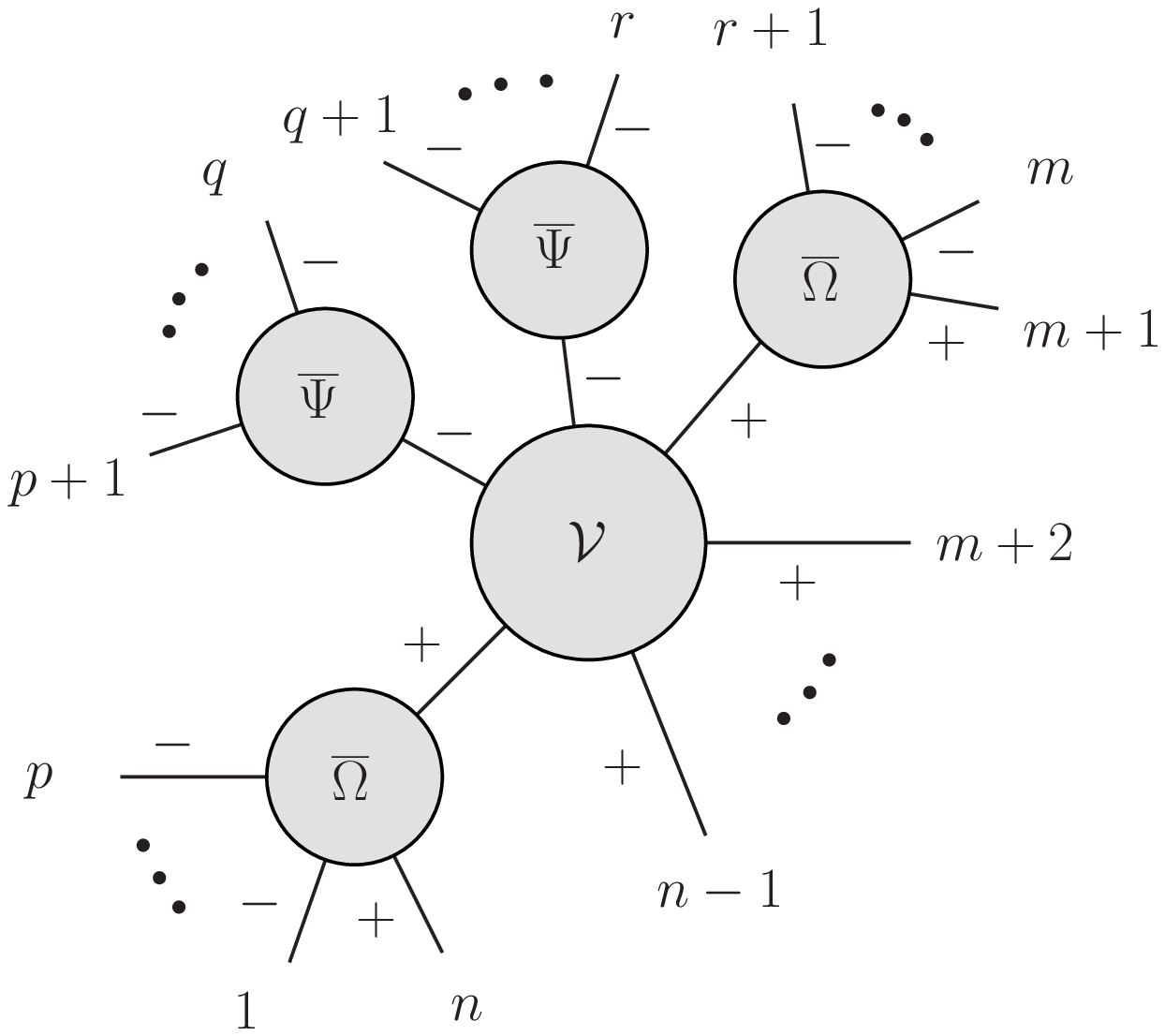}}
    \caption{\small Left: color ordered vertex in the $Z$-field theory with $m$ minus helicity legs. Right: a general contribution to the $Z$-theory vertex. The central blob is the MHV vertex.
    }
    \label{fig:vertex}
\end{figure}

We observe, that the most general contribution has the form  depicted in Fig.~\ref{fig:vertex}. 
Indeed, substitution of the plus helicity field $B^{\bullet}$ in the MHV vertex by the $Z$ fields, Eq.~\eqref{eq:BbulletZ_exp}, results in one plus helicity leg in addition to several negative helicity legs.
Thus it must be adjacent to other minus helicity legs multiplicated
by the $\overline{\Psi}$ kernels. Therefore, there can be at most two $\overline{\Omega}$ kernels. Also, there can be at most two $\overline{\Psi}$ kernels, because there are only two negative helicity legs in the MHV vertex. In addition to the above general situation, there are cases when the kernels are trivial, i.e.  $\overline{\Psi}_1=1$ or  $\overline{\Omega}_1=1$. 
Let us note, that the MHV vertex has always outgoing momenta, but the 
$\overline{\Psi}$ and  $\overline{\Omega}$ kernels have the off-shell line incoming. In addition, there is \emph{no} propagator connecting the kernels with the MHV vertex. Therefore, we can identify the internal lines by a single helicity flow, given by the helicity of the MHV vertex.

Specifically, consider now the vertex for $n$ external legs with the momenta $\mathbf{p}_1\dots \mathbf{p}_n$, where $\mathbf{p}_1\dots \mathbf{p}_m$ correspond to the minus helicity legs. Let us introduce a collective  index $[i,i+1,\dots,j]$ labeling the momentum, $\mathbf{p}_{i(i+1)\dots j}=\mathbf{p}_i+\mathbf{p}_{i+1}+\dots+\mathbf{p}_j$. Using this notation, the general form of the color ordered vertex can be written as: 
\begin{multline}
    \mathcal{U}\left(1^-,2^-,\dots,m^-,(m\!+\!1)^+,\dots,n^+\right) = 
    \sum_{p=0}^{m-2}\sum_{q=p+1}^{m-1}\sum_{r=q+1}^{m}\\
    \mathcal{V}\left(\,[p\!+\!1,\dots,q]^-,[q\!+\!1,\dots,r]^-,[r\!+\!1,\dots,m\!+\!1]^+,(m\!+\!2)^+,\dots,(n\!-\!1)^+,[n,1,\dots,p]^+\right) \\
    \overline{ \Omega}\left(n^+,1^-,\dots,p^-\right) \,\,
    \overline{ \Psi}\left((p\!+\!1)^-,\dots,q^-\right) \,\, 
    \overline{ \Psi}\left((q\!+\!1)^-,\dots,r^-\right) \,\, 
    \overline{ \Omega}\left((r\!+\!1)^-,\dots,m^-,(m\!+\!1)^+\right) \,\,
    \label{eq:Z_gen_ker}
\end{multline}
Although the analytic formulas do not seem to collapse, in general, to any simple form, the above  expression is operational and
 can be readily applied in the actual amplitude calculation, as we shall demonstrate in Section~\ref{sec:Applications}.

%-----------------------------------------------
\subsection{Summary: Feynman rules for the new action}

Let us summarize the content of the new action. 
It contains a set of vertices with increasing multiplicity, starting at $n=4$. Each vertex has at least two minus helicity legs, and at most $n-2$. Thus, we have MHV vertices, next-to-MHV (NMHV) vertices, next-to-next-to-MHV (NNMHV) vertices and so on. The vertices with maximal number of minus helicity legs are just $\overline{\text{MHV}}$ vertices. Both the MHV and $\overline{\text{MHV}}$ vertices alone give the corresponding on-shell amplitudes. In addition to these vertices we have a scalar propagator joining two opposite helicity legs.

In the following section we shall demonstrate how various amplitudes are calculated. To this end we introduce the following color-ordered Feynman rules:
\begin{enumerate}[label={\it\roman*}$\,$)]
\item \emph{scalar propagator} joining a plus and a minus helicity leg \\
\begin{center}
\includegraphics[width=4cm]{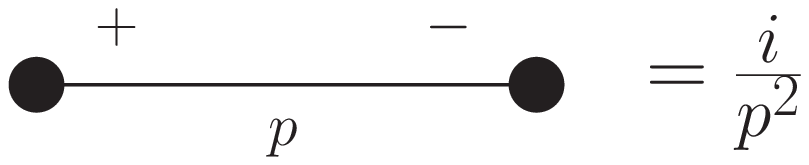}
\end{center}
\item \emph{$n$-point vertex}, $n\geq4$, with $m$ negative helicity legs, $2\leq m \leq n-2$ \\
\begin{center}
\includegraphics[width=10cm]{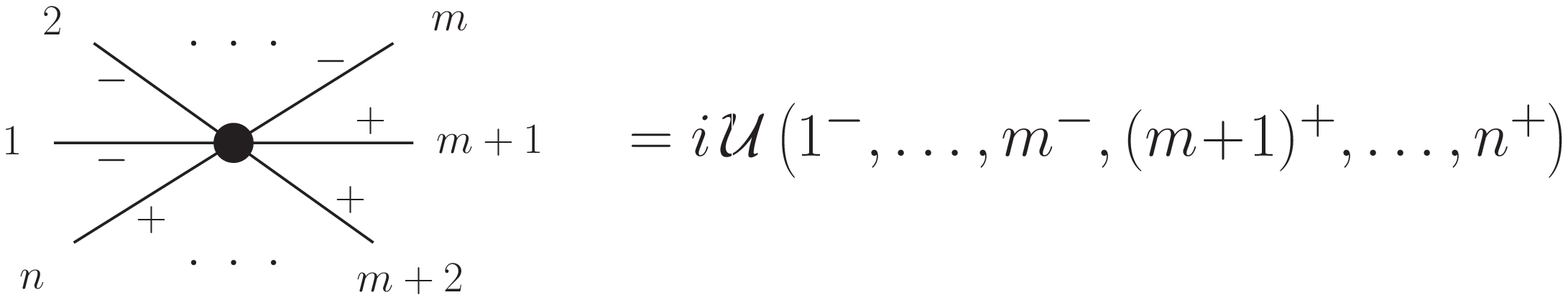}
\end{center}
\end{enumerate}

%-----------------------------------------------
%-----------------------------------------------
\section{Applications}
\label{sec:Applications}

In this section we shall calculate several tree amplitudes using the new action. 

%-----------------------------------------------
\subsection{4-point and 5-point amplitudes}

The lowest non-zero amplitude is the 4-point MHV amplitude. It is simply given by the 4-point vertex in the theory.

The non-zero 5-point amplitudes are MHV $(--+++)$ and $\overline{\text{MHV}}$ $(---++)$. Both amplitudes are given just by a single vertex, respectively $\mathcal{U}(1^-,2^-,3^+,4^+,5^+)$ and $\mathcal{U}(1^-,2^-,3^-,4^+,5^+)$. For the MHV it is the expression \eqref{eq:MHV_vertex_colororder}, giving for the amplitude
\begin{equation}
    \mathcal{A}(1^-,2^-,3^+,4^+,5^+) =
 -g^3 \left(\frac{p_{1} ^{+}}{p_{2}^{+}}\right)^{2}
\frac{\widetilde{v}_{21}^{*4}}{\widetilde{v}^*_{15}\widetilde{v}^*_{54}\widetilde{v}^*_{43}  \widetilde{v}^*_{32}\widetilde{v}^*_{21} } \, .
 \label{eq:5G_MHV_onshell}
\end{equation}

\begin{figure}
    \centering
 \includegraphics[width=13cm]{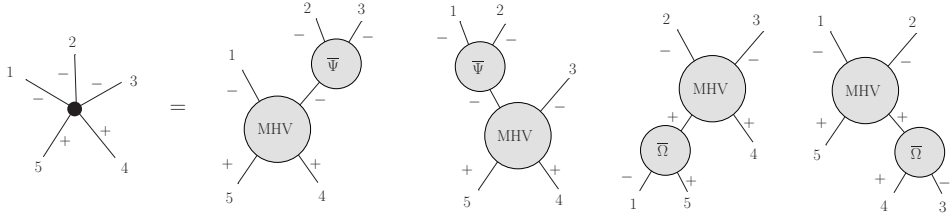}
    \caption{\small 
    The contributions to the color-ordered $\overline{\text{MHV}}$ vertex, with helicity $(---++)$.
    }
    \label{fig:MHVbar5_vertex}
\end{figure}

For the $\overline{\text{MHV}}$  the expression is easily obtained from \eqref{eq:Z_gen_ker}. In Fig.~\ref{fig:MHVbar5_vertex} we show the contributing terms.
Using the explicit expressions we have: 
\begin{multline}
 \mathcal{U}(1^-,2^-,3^-,4^+,5^+) = g^3  \Bigg[  \left(\frac{p_{1} ^{+}}{p_{23}^{+}}\right)^{2}
\frac{\widetilde{v}_{(23)1}^{*4}}{\widetilde{v}_{15}^{*}\widetilde{v}_{54}^{*}\widetilde{v}_{4(23)}^{*}  \widetilde{v}_{({23})1}^{*} } \times  \frac{{\widetilde v}_{({23})2}}{{\widetilde v}_{32}{\widetilde v}_{2({23})}}   \\
+     \left(\frac{p_{12} ^{+}}{p_{3}^{+}}\right)^{2}
\frac{\widetilde{v}_{3({12})}^{*4}}{\widetilde{v}_{({12}){5}}^{*}\widetilde{v}_{54}^{*}\widetilde{v}_{43}^{*}  \widetilde{v}_{{3}({12})}^{*} } \times  \frac{{\widetilde v}_{({12})1}}{{\widetilde v}_{21}{\widetilde v}_{1({12})}}\\
+ \left(\frac{p_{2} ^{+}}{p_{3}^{+}}\right)^{2}
\frac{\widetilde{v}_{32}^{*4}}{\widetilde{v}_{2({15})}^{*}\widetilde{v}_{({15})4}^{*}\widetilde{v}_{43}^{*}  \widetilde{v}_{32}^{*} } \times \left(\frac{p_{5} ^{+}}{p_{15}^{+}}\right)^{2} \frac{{\widetilde v}_{({15})5}}{{\widetilde v}_{15}{\widetilde v}_{5({15})}}    \\ 
+    \left(\frac{p_{1} ^{+}}{p_{2}^{+}}\right)^{2}
\frac{\widetilde{v}_{21}^{*4}}{\widetilde{v}_{15}^{*}\widetilde{v}_{5\left(34\right)}^{*}\widetilde{v}_{({34)}2}^{*}  \widetilde{v}_{21}^{*} } \times \left(\frac{p_{4} ^{+}}{p_{34}^{+}}\right)^{2} \frac{{\widetilde v}_{({34})3}}{{\widetilde v}_{43}{\widetilde v}_{3({34})}}  \Bigg]\, .
 \label{eq:5g_MHVbar}
\end{multline}
We have checked, that the above expression reduces in the on-shell limit to the known formula for the $\overline{\text{MHV}}$ amplitude:
\begin{equation}
    \mathcal{A}(1^-,2^-,3^-,4^+,5^+) =
 g^3 \left(\frac{p_{4} ^{+}}{p_{5}^{+}}\right)^{2}
\frac{\widetilde{v}_{54}^{4}}{\widetilde{v}_{15}\widetilde{v}_{54}\widetilde{v}_{43}  \widetilde{v}_{32}\widetilde{v}_{21} } \, .
 \label{eq:5G_MHVbar_onshell}
\end{equation}

%-----------------------------------------------
\subsection{6-point amplitudes}

The MHV and $\overline{\text{MHV}}$ amplitudes are always given by the single vertices. For the latter we need the $\mathcal{U}(1^-,2^-,3^-,4^-,5^+,6^+)$ vertex which is given by the formula \eqref{eq:Z_gen_ker}, see Appendix~\ref{sec:AppB}. We have verified that it recovers the correct result in the on-shell limit.

\begin{figure}[h]
    \centering
 \includegraphics[width=13cm]{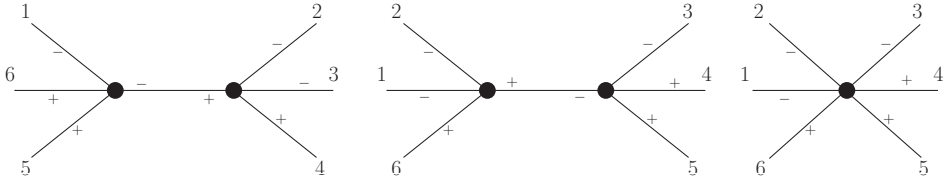}
    \caption{\small 
    Diagrams contributing to the 6-point NMHV amplitude $(---+++)$.
    }
    \label{fig:NMHV6}
\end{figure}

The remaining amplitude is the  NMHV amplitude with helicity configuration $(---+++)$. We have just three contributing diagrams depicted in Fig.~\ref{fig:NMHV6}. First two diagrams connect two MHV vertices, whereas the last one is the NMHV vertex, given by $\mathcal{U}(1^-,2^-,3^-,4^+,5^+,6^+)$.
We have checked that the sum of those diagrams reproduce  the known result in the on-shell limit \cite{Kosower1990}.

%-----------------------------------------------
\subsection{7-point amplitudes}

In addition to the MHV and $\overline{\text{MHV}}$ amplitudes, which are again calculated through just a single vertex, we have the NMHV and NNMHV amplitudes.

\begin{figure}[h]
    \centering
 \includegraphics[width=13cm]{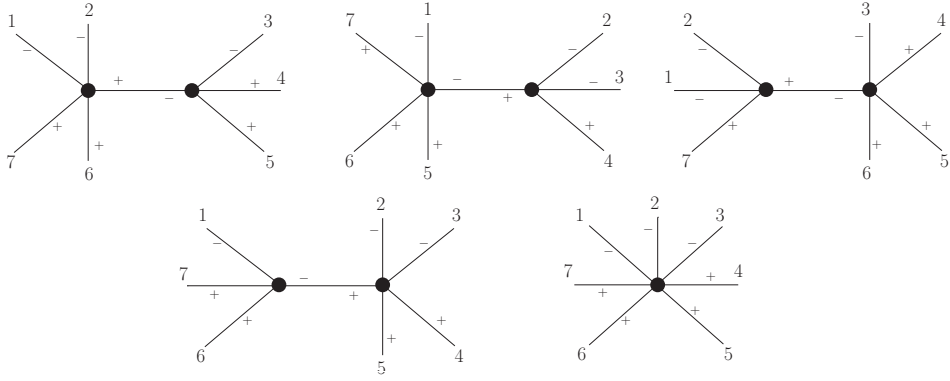}
    \caption{\small 
    Diagrams contributing to the 7-point NMHV amplitude $(---++++)$.
    }
    \label{fig:NMHV7}
\end{figure}

The diagrams contributing to the NMHV amplitude are depicted in Fig.~\ref{fig:NMHV7}. We have four diagrams connecting 4-point and 5-point MHV vertices and one diagram which consists of the single vertex with NMHV helicity configuration. The corresponding expression is
\begin{equation}
    D_1+D_2+D_3+D_4+\frac{1}{2}D_5 \,,
    \label{eq:sum_d5}
\end{equation}
where the individual diagrams are calculated as:
\begin{align}
    D_1&=i\, \mathcal{U} \left(1^-,2^-,[3,4,5]^+,6^+,7^+\right) \times \frac{i}{p^2_{6712}}\times i\,\mathcal{U} \left([6,7,1,2]^-,3^-,4^+,5^+\right) \, \\
    D_2&= i\,\mathcal{U} \left(1^-,[2,3,4]^-,5^+,6^+,7^+\right) \times \frac{i}{p^2_{234}}\times i\,\mathcal{U} \left(2^-,3^-,4^+,[5,6,7,1]^+\right) \, \\
    D_3&=i\, \mathcal{U} \left(1^-,2^-,[3,4,5,6]^+,7^+\right) \times \frac{i}{p^2_{3456}}\times i\,\mathcal{U} \left([7,1,2]^-,3^-,4^+,5^+,6^+\right) \, \\
    D_4&= i\,\mathcal{U} \left(1^-,[2,3,4,5]^-,6^+,7^+\right) \times \frac{i}{p^2_{2345}}\times i\,\mathcal{U}\left(2^-,3^-,4^+,5^+,[6,7,1]^+\right) \\
    D_5&= i\,\mathcal{U} \left(1^-,2^-,3^-,4^+,5^+,6^+,7^+\right)
    \,.
    \label{eq:7nmhv_z}
\end{align}
In the  above, $i/p^2_{i_1\dots i_m}=i/(p_{i_1}+\dots+p_{i_m})^2$ is the scalar propagator. 
The factor $1/2$ multiplying $D_5$ comes from the fact that there are two possible color orders contributing  to diagrams $D_1$-$D_4$.
We have compared the sum of those diagrams, i.e. Eq.\eqref{eq:sum_d5} with the on-shell result obtained using the \texttt{GGT} Mathematica package \cite{Dixon:2010ik} together with the \texttt{S@M} package \cite{Maitre2007}  and found an exact match, up to an overall normalization due to the difference in our symbol $\tilde{v}_{ij}$ and $\left<ij\right>$.

\begin{figure}[h]
    \centering
 \includegraphics[width=13cm]{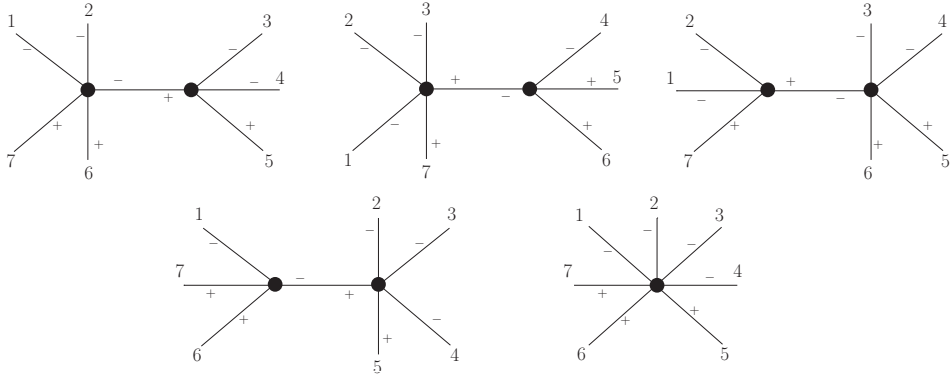}
    \caption{\small 
    Diagrams  contributing to the 7-point NNMHV amplitude $(----+++)$.
    }
    \label{fig:NNMHV7}
\end{figure}

For the NNMHV amplitude the number of diagrams stays the same, see Fig.~\ref{fig:NNMHV7}. Here, however, we encounter a new feature, namely there are now diagrams that connect the 4-point MHV vertex with 5-point $\overline{\text{MHV}}$ vertex. That is, starting with this amplitude we utilize the new vertices appearing in the theory in a nontrivial way (i.e. by gluing them with other vertices). We again find the on shell limit of the result consistent with the \texttt{GGT} package.

%-----------------------------------------------
\subsection{8-point amplitudes}

We have also calculated some of the non-trivial 8-point amplitudes, all of which agree with the standard results calculated numerically using the \texttt{GGT} package.

\begin{figure}[h]
    \centering
 \includegraphics[width=13cm]{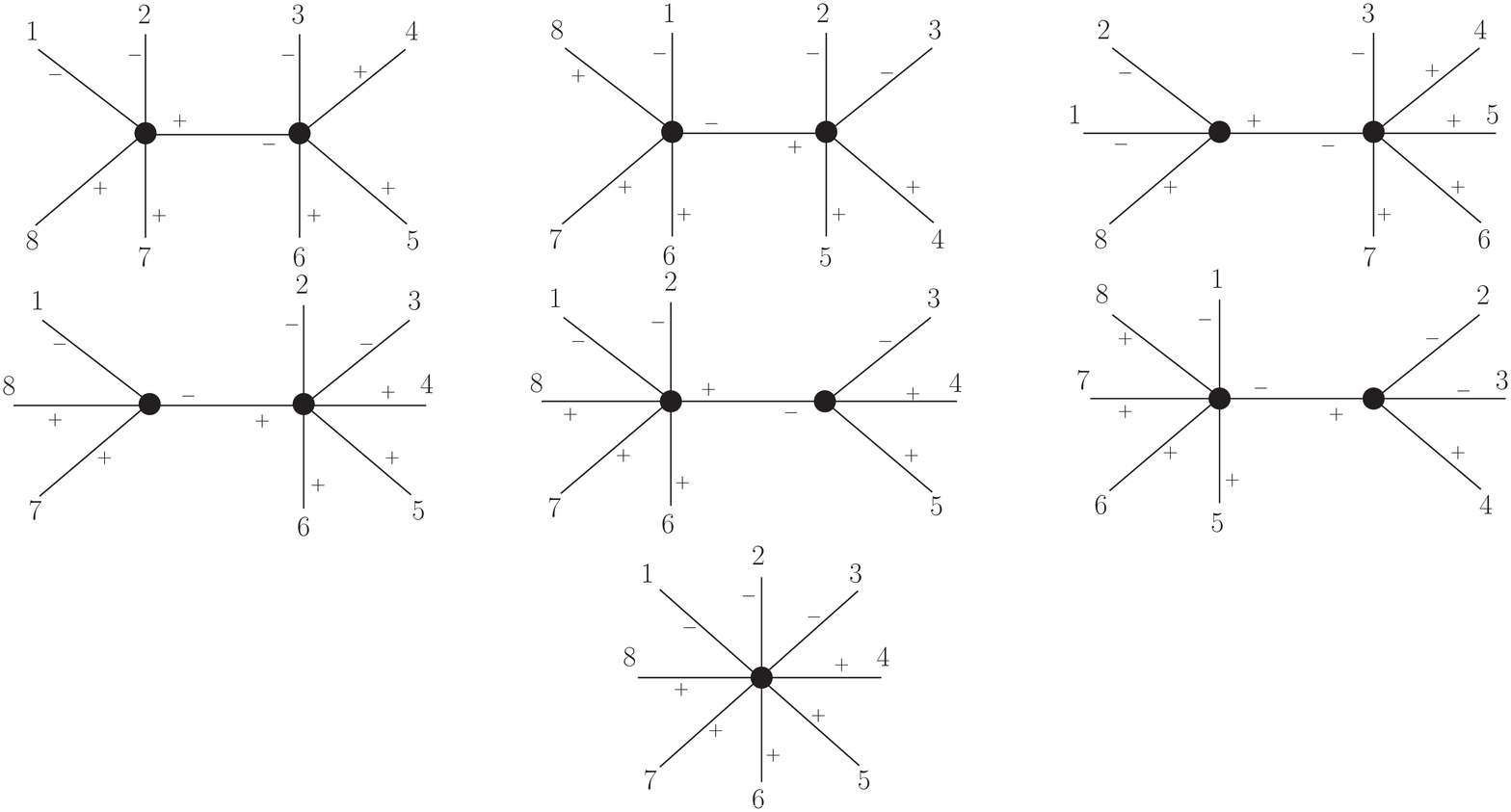}
    \caption{\small 
    Diagrams  contributing to the 8-point NMHV amplitude $(---+++++)$.
    }
    \label{fig:NMHV8}
\end{figure}

The 8-point NMHV amplitude turns out to be actually very simple, requiring only 7 diagrams, of which 6 join two MHV vertices, see Fig.~\ref{fig:NMHV8}.

\begin{figure}[h]
    \centering
 \includegraphics[width=13cm]{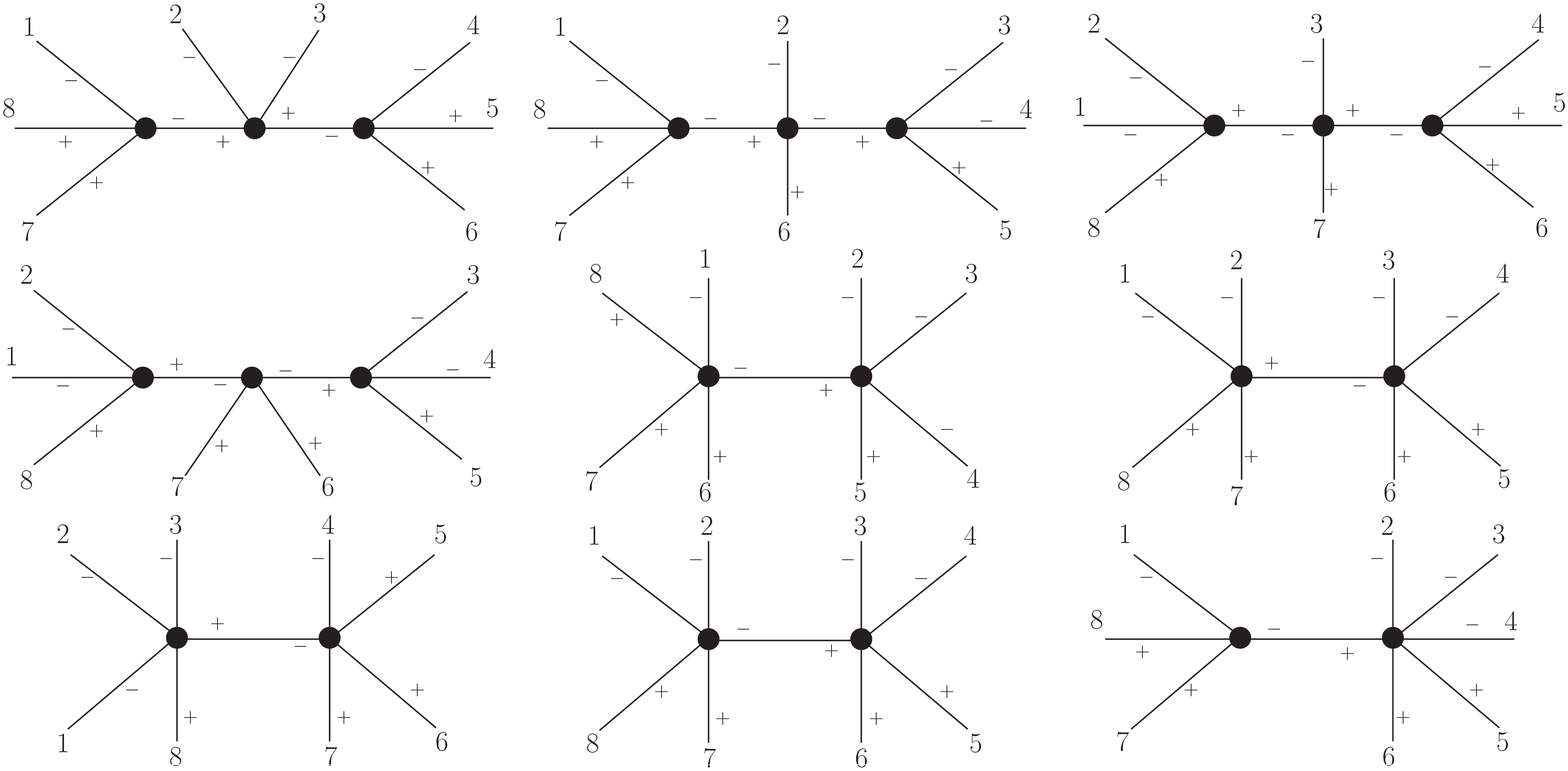}\\
 \includegraphics[width=13cm]{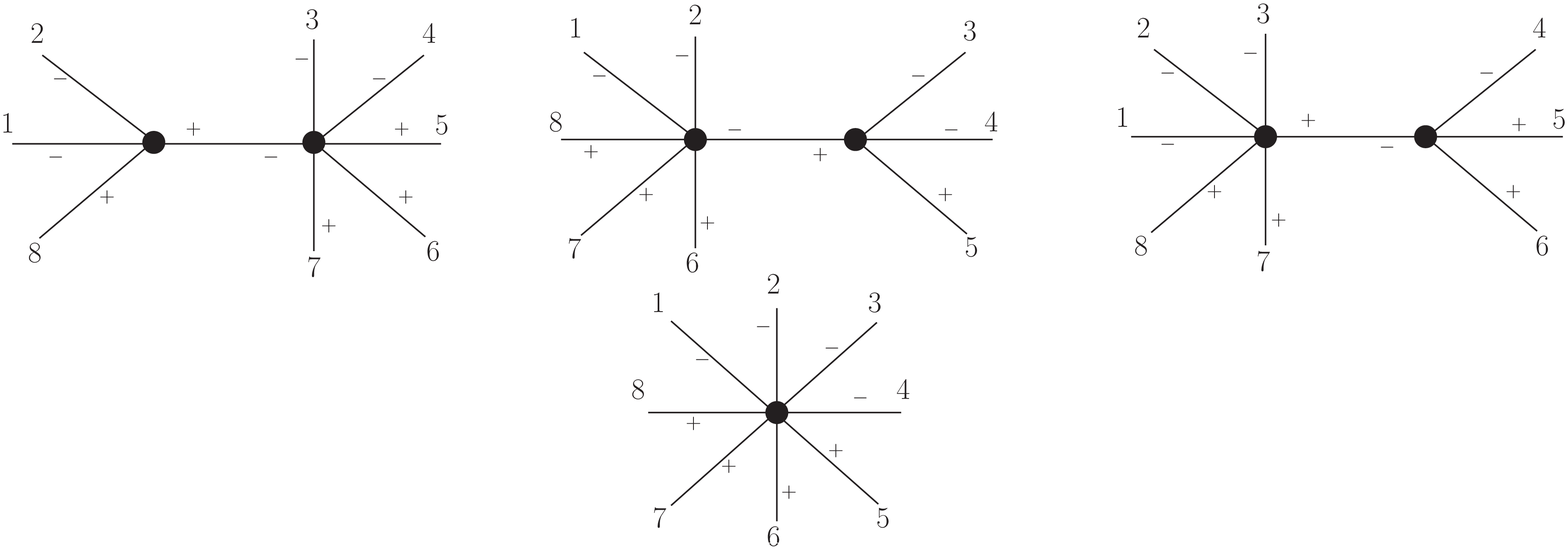}
    \caption{\small 
    Diagrams contributing to the 8-point NNMHV amplitude $(----++++)$.
    }
    \label{fig:NNMHV8}
\end{figure}

In order to really test the new theory we calculated the 8-point NNMHV amplitude. In that case, we encounter only 13 diagrams, shown in Fig.~\ref{fig:NNMHV8}, of which some consist of three MHV vertices joined by the scalar propagators, but there are also diagrams combining the 5-point $\overline{\text{MHV}}$ vertex as well as 6-point NMHV vertex appearing in the $Z$-field theory.
As this calculation is less trivial, let us list the structure of the diagrams (they correspond to the diagrams in Fig.~\ref{fig:NNMHV8} from the top left, going to the right):
\begin{multline}
    D_1 = i\,\mathcal{U} \left([7,8,1,2,3]^-, 4^-,5^+, 6^+\right) \times \frac{i}{p^2_{456}}\times i\,\mathcal{U} \left(1^-, [2,3,4,5,6]^-,7^+,8^+\right) \\ 
    \times \frac{i}{p^2_{781}} 
    \times i\,\mathcal{U} \left(2^-, 3^-, [4,5,6]^+,[7,8,1]^+\right) \, ,
\end{multline}
\vspace{-0.2cm}
\begin{multline}
    D_2 = i\,\mathcal{U} \left(3^-,4^-,5^+,[6,7,8,1,2]^+\right) \times \frac{i}{p^2_{67812}}\times i\,\mathcal{U} \left(1^-, [2,3,4,5,6]^-,7^+,8^+\right) \\
     \times \frac{i}{p^2_{781}}\times i\,\mathcal{U} \left(2^-, [3,4,5,]^-, 6^+,[7,8,1]^+\right) \, ,
\end{multline}   
\vspace{-0.2cm}
\begin{multline}
    D_3= i\,\mathcal{U} \left([7,8,1,2,3]^-,4^-,5^+, 6^+\right\} \times \frac{i}{p^2_{456}}\times i\,\mathcal{U} \left(1^-, 2^-, [3,4,5,6,7]^+,8^+\right)\\
     \times \frac{i}{p^2_{812}}\times i\,\mathcal{U} \left([8,1,2]^-,3^-,[4,5,6]^+, 7^+\right) \,, 
\end{multline}   
\vspace{-0.2cm}
\begin{multline}    
    D_4= i\,\mathcal{U} \left(3^-,4^-,5^+,[6,7,8,1,2]^+\right) \times \frac{i}{p^2_{67812}}\times i\,\mathcal{U} \left(1^-, 2^-, [3,4,5,6,7]^+,8^+\right)\\  \times \frac{i}{p^2_{812}}\times i\,\mathcal{U} \left([8,1,2]^-,[3,4,5]^-,6^+, 7^+\right) \,,
\end{multline}    
\begin{equation}    
    D_5= i\,\mathcal{U} \left(1^-,[2,3,4,5]^-,6^+,7^+, 8^+\right) \times \frac{i}{p^2_{2345}}\times i\,\mathcal{U} \left(2^-, 3^-,4^-,5^+,[6,7,8,1]\right)\,,
\end{equation}    
\begin{equation}    
    D_6= i\,\mathcal{U} \left(1^-,2^-,[3,4,5,6]^+,7^+, 8^+\right) \times \frac{i}{p^2_{3456}}\times i\,\mathcal{U} \left([7,8,1,2]^-,3^-,4^-,5^+,6^+\right)\,,
\end{equation}    
\begin{equation}    
    D_7= i\,\mathcal{U} \left(1^-,2^-,3^-,[4,5,6,7]^+,8^+\right) \times \frac{i}{p^2_{4567}}\times i\,\mathcal{U} \left([8,1,2,3]^-,4^-,5^+,6^+,7^+\right)\, ,
\end{equation}    
\begin{equation}    
    D_8= i\,\mathcal{U} \left(1^-,2^-,[3,4,5,6]^-,7^+,8^+\right) \times \frac{i}{P^2_{3456}}\times i\,\mathcal{U} \left(3^-, 4^-,5^+,6^+,[7,8,1,2]^+\right)\,,
\end{equation}    
\begin{equation}    
    D_9= i\,\mathcal{U} \left(1^-,[2,3,4,5,6]^-,7^+,8^+\right) \times \frac{i}{p^2_{23456}}\times i\,\mathcal{U} \left(2^-, 3^-,4^-,5^+,6^+,[7,8,1]^+\right)\,, 
\end{equation}   
\begin{equation}    
    D_{10}= i\,\mathcal{U} \left(1^-,2^-,[3,4,5,6,7]^+, 8^+\right\} \times \frac{i}{p^2_{34567}}\times i\,\mathcal{U} \left([8,1,2]^-,3^-,4^-,5^+,6^+, 7^+\right)\,, 
\end{equation}   
\begin{equation}    
    D_{11}= i\,\mathcal{U} \left(3^-,4^-,5^+,[6,7,8,1,2]^+\right) \times \frac{i}{p^2_{67812}}\times i\,\mathcal{U} \left(1^-, 2^-,[3,4,5]^-,6^+,7^+,8^+\right)\,,
\end{equation}    
\begin{equation}    
    D_{12}= i\,\mathcal{U} \left([7,8,1,2,3]^-,4^-,5^+, 6^+\right) \times \frac{i}{p^2_{456}}\times i\,\mathcal{U} \left(1^-,2^-,3^-,[4,5,6]^+,7^+, 8^+\right)\, , 
\end{equation}    
\begin{equation}    
    D_{13}= i\,\mathcal{U} \left(1^-,2^-,3^-,4^-,5^+,6^+,7^+,8^+\right) \, .
\end{equation}
The above diagrams have to be combined as follows, due to the additional combinatorial factors as explained in the previous subsection:
\begin{equation}
    D_1+D_2+D_3+D_4+\frac{1}{2}\left(D_5+D_6+D_7+D_8+D_9+D_{10}+D_{11}+D_{12}\right) + \frac{1}{4} D_{13} \,.
\end{equation}

%-----------------------------------------------
%-----------------------------------------------
\section{Discussion}
\label{sec:Summary}

\begin{table}
\centering
\renewcommand{\arraystretch}{1.4}
\begin{tabular}{cccc}
\cline{1-3}
\multicolumn{1}{|c|}{\# legs}             & \multicolumn{1}{c|}{helicity} & \multicolumn{1}{c|}{\# diagrams} &  \\ %\cline{1-3}
\hhline{===}
\multicolumn{1}{|c|}{\multirow{2}{*}{4 point}} & \multicolumn{1}{c|}{$\mathrm{MHV}$}           & \multicolumn{1}{c|}{1}                                                                         &  \\ \cline{2-3}
\multicolumn{1}{|c|}{}                         & \multicolumn{1}{c|}{$\mathrm{\overline{MHV}}$}           & \multicolumn{1}{c|}{1}                                                                         &  \\ \cline{1-3}
\multicolumn{1}{|c|}{\multirow{2}{*}{5 point}} & \multicolumn{1}{c|}{$\mathrm{MHV}$}           & \multicolumn{1}{c|}{1}                                                                         &  \\ \cline{2-3}
\multicolumn{1}{|c|}{}                         & \multicolumn{1}{c|}{$\mathrm{\overline{MHV}}$}           & \multicolumn{1}{c|}{1}                                                                         &  \\ \cline{1-3}
\multicolumn{1}{|c|}{\multirow{3}{*}{6 point}} & \multicolumn{1}{c|}{$\mathrm{MHV}$}           & \multicolumn{1}{c|}{1}                                                                         &  \\ \cline{2-3}
\multicolumn{1}{|c|}{}                         & \multicolumn{1}{c|}{$\mathrm{NMHV}$}           & \multicolumn{1}{c|}{3}                                                                         &  \\ \cline{2-3}
\multicolumn{1}{|c|}{}                         & \multicolumn{1}{c|}{$\mathrm{\overline{MHV}}$}           & \multicolumn{1}{c|}{1}                                                                         &  \\ \cline{1-3}
\multicolumn{1}{|c|}{\multirow{4}{*}{7 point}} & \multicolumn{1}{c|}{$\mathrm{MHV}$}           & \multicolumn{1}{c|}{1}                                                                         &  \\ \cline{2-3}
\multicolumn{1}{|c|}{}                         & \multicolumn{1}{c|}{$\mathrm{NMHV}$}           & \multicolumn{1}{c|}{5}                                                                         &  \\ \cline{2-3}
\multicolumn{1}{|c|}{}                         & \multicolumn{1}{c|}{$\mathrm{NNMHV}$}           & \multicolumn{1}{c|}{5}                                                                         &  \\ \cline{2-3}
\multicolumn{1}{|c|}{}                         & \multicolumn{1}{c|}{$\mathrm{\overline{MHV}}$}           & \multicolumn{1}{c|}{1}                                                                         &  \\ \cline{1-3}
\multicolumn{1}{|c|}{\multirow{4}{*}{8 point}} & \multicolumn{1}{c|}{$\mathrm{MHV}$}           & \multicolumn{1}{c|}{1}                                                                         &  \\ \cline{2-3}
\multicolumn{1}{|c|}{}                         & \multicolumn{1}{c|}{$\mathrm{NMHV}$}           & \multicolumn{1}{c|}{7}                                                                         &  \\ \cline{2-3}
\multicolumn{1}{|c|}{}                         & \multicolumn{1}{c|}{$\mathrm{NNMHV}$}           & \multicolumn{1}{c|}{13}                                                                         &  \\ \cline{2-3}
\multicolumn{1}{|c|}{}                         & \multicolumn{1}{c|}{$\mathrm{NNNMHV}$}           & \multicolumn{1}{c|}{7}                                                                        &  \\ \cline{2-3}
\multicolumn{1}{|c|}{}                         & \multicolumn{1}{c|}{$\mathrm{\overline{MHV}}$}           & \multicolumn{1}{c|}{1}                                                                         &  \\ \cline{1-3}
                                               &                                 &                                                                                                & 
\end{tabular}
\caption{\small
Total number of diagrams for helicity amplitudes of different multiplicities.}
\label{table:Z_th_ampl}
\end{table}

In the present work we have constructed a new action for gluodynamics  by applying two consecutive canonical transformations on the light-cone Yang-Mills action, see Fig.~\ref{fig:CT_paths}. The same action can be also obtained by a single canonical transformation, whose generating functional is given by Eq.~\eqref{eq:generatingfunc2}. 

The most striking property of the new action is that it has no triple-gluon vertex.
Effectively, the triple-gluon vertices are resummed inside the Wilson lines (see \cite{Kotko2017} for the explicit demonstration of that fact for the $(++-)$ vertex). 
Consequently, the number of diagrams needed to calculate the amplitudes is reduced, as compared for example to the CSW method.  
 For example, for the NMHV amplitudes with adjacent helicity, the CSW rules give $2(n-3)$ diagrams, whereas the new theory gives $2(n-5)+1$ diagrams, $n\geq 5$.
We give the number of diagrams for various adjacent helicity configurations in Table~\ref{table:Z_th_ampl}.  It is important to stress, that we do not mean here the number of contributing terms, as the vertices in the new theory are not, in general, given by a single term in our representation. The structure of those vertices can be however easily obtained by means of the master equation \eqref{eq:Z_gen_ker}.

\begin{figure}
    \centering
    \includegraphics[width=11cm]{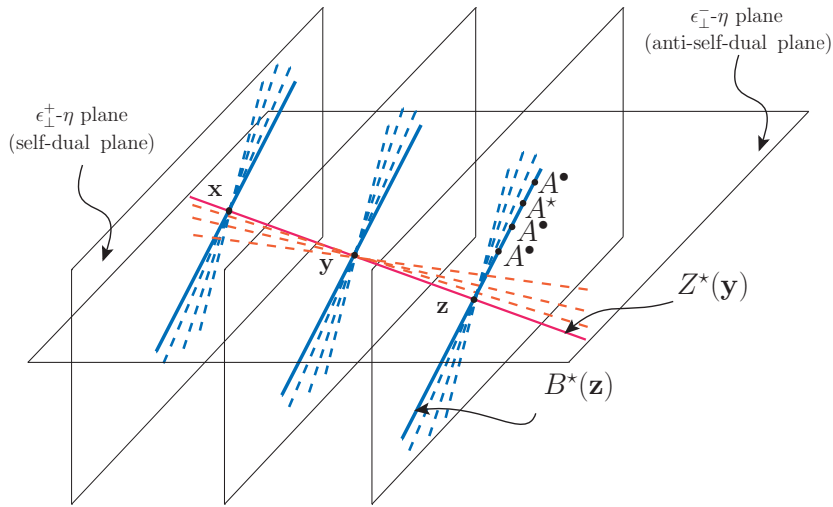}
    \caption{\small
    Schematic presentation of the geometric structure of the $Z^{\star}$ field (the structure of $Z^{\bullet}$ is quite similar). The vertical planes are the self-dual planes, i.e. the planes spanned by the vectors $\varepsilon_{\perp}^+=(0,1,i,0)$ and $\eta=(1,0,0,-1)$. The horizontal plane, in our terminology, is the anti-self-dual plane, i.e. it is spanned by $\varepsilon_{\perp}^-=(0,1,-i,0)$ and $\eta$.
    The $B^{\star}$ fields (the minus helicity fields in the MHV action) are the straight infinite Wilson lines on the self-dual plane, integrated over all slopes, and differentiated functionally to replace one plus gluon helicity field by the minus gluon helicity field. This structure is represented by the blue lines.
    The $Z^{\star}$ field, i.e. the minus helicity field in the new theory, is  given by a similar Wilson line of the $B^{\star}$ fields, lying on the anti-self dual plane, and integrated over all slopes. 
    }
    \label{fig:geometry}
\end{figure}

One of the very interesting aspects of the field transformations leading to the new action is its incredibly rich geometric structure. Let us recall, that the new minus helicity field $Z^{\star}$ is given by the straight infinite Wilson line on the anti-self-dual plane (i.e. the plane spanned by $\varepsilon_{\perp}^-$ and $\eta$), integrated over all directions.  This is the Wilson line functional of the minus helicity field of the MHV theory, which itself is given as the analogous Wilson line of the usual gauge fields lying on the self-dual plane, with insertion of the minus helicity field (i.e. the functional derivative, see Eq.~\eqref{eq:Zfield_transform}). We schematically depict the structure of the $Z^{\star}$ field in Fig.~\ref{fig:geometry}. A similar figure can be drawn for the $Z^{\bullet}$ field. We stress, that although the overall picture looks very non-local, it is local in the light-cone time. 

In our recent work \cite{Kakkad2020} we have discussed the fields in the MHV theory, where the fields can be expressed as the Wilson line functionals of the standard gauge fields lying exclusively on the self-dual plane. We see that the transformations derived in the present work extend this picture to the whole 3-space, with the light-cone time fixed.

One of the future directions is obviously to find the quantum corrections to the new action. There is an immediate difficulty in that program, namely, the fact that in the quantum MHV action there are contributions evading the S-matrix equivalence theorem \cite{Ettle2007}, and we expect similar contributions in the new action. An alternative approach is based on the world-sheet regularization, which successfully recovered one loop all-plus helicity vertex in the MHV action \cite{Brandhuber2007}.

Another interesting direction of future study is related to the rich geometric structure of the transformations, sketched in Fig.~\ref{fig:geometry}, which has not been fully explored.

%-----------------------------------------------
%-----------------------------------------------
\section{Acknowledgments}
\label{sec:Acknowledgments}
H.K. and P.K. are supported by the National Science Centre, Poland grant no. 2018/31/D/ST2/02731.  A.M.S. is supported  by the U.S. Department of Energy Grant 
 DE-SC-0002145 and  in part by  National Science Centre in Poland, grant 2019/33/B/ST2/02588.
%-----------------------------------------------
%-----------------------------------------------

\bibliographystyle{JHEP}
\bibliography{library}

%-----------------------------------------------%-----------------------------------------------

\newpage
\appendix

%-----------------------------------------------
%-----------------------------------------------
\section{Cancellation of triple-gluon vertices.}
\label{sec:AppA}

In this appendix we show the cancellation of both the triple-gluon vertices when transforming the Yang-Mills fields to the new fields,
\begin{equation}
    \left\{\hat{A}^{\bullet},\hat{A}^{\star}\right\} \rightarrow \Big\{\hat{Z}^{\bullet}\big[{A}^{\bullet},{A}^{\star}\big],\hat{Z}^{\star}\big[{A}^{\bullet},{A}^{\star}\big]\Big\} \, .
    \label{eq:general_transf_app}
\end{equation}

As previously argued, there are two ways of doing this: using directly the generating functional \eqref{eq:generatingfunc3}, or using two consecutive canonical field transformations. For convenience, we shall follow the second path. So the strategy is to express $A$ fields in terms of $B$ fields and then substitute for $B$ fields in terms of $Z$ fields. Then we shall substitute the $A$ fields (expressed already in terms of $Z$ fields up to the second order) into the standard Yang-Mills action \eqref{eq:YM_LC_action} to show that the triple-gluon vertices cancel out.

Using the relations \eqref{eq:B_bull_exp}-\eqref{eq:B_star_exp} and \eqref{eq:BstarZ_exp}-\eqref{eq:BbulletZ_exp}, it is easy to see that the expansion of $\widetilde{A}^{\bullet}$, $\widetilde{A}^{\star}$ fields, to second order in $\widetilde{Z}$ fields is the following:
\begin{multline}
    \Big[{\widetilde {A}}^{\bullet}_{a} (x^+;\mathbf{P})\Big]_{2nd} =\int\!d^{3}\mathbf{p}_{1} d^{3}\mathbf{p}_{2}\, \overline{\widetilde \Omega}\,^{a b_1 \left \{b_2 \right \}}_{2}(\mathbf{P}; \mathbf{p_1} ,\left \{ \mathbf{p_2} \right \}) {\widetilde Z}^{\bullet}_{b_1} (x^+;\mathbf{p}_{1}) {\widetilde Z}^{\star}_{b_2} (x^+;\mathbf{p}_{2}) \\
    + \int\!d^{3}\mathbf{p}_{1}\, d^{3}\mathbf{p}_{2} {\widetilde \Psi}_{2}^{a \left \{b_1 b_2 \right \} }(\mathbf{P}; \left \{\mathbf{p}_{1}, \mathbf{p}_{2} \right \}) {\widetilde Z}^{\bullet}_{b_1} (x^+;\mathbf{p}_{1}){\widetilde Z}^{\bullet}_{b_2} (x^+;\mathbf{p}_{2}) \, ,
    \label{eq:A_bull_2}
\end{multline}
and
\begin{multline}
    \Big[{\widetilde A}^{\star}_{a} (x^+;\mathbf{P})\Big]_{2nd} = \int\!d^{3}\mathbf{p}_{1}\, d^{3}\mathbf{p}_{2}\, \overline{\Psi}\,^{a\{b_1 b_2\}}_2(\mathbf{P};\{\mathbf{p}_1,\mathbf{p}_2\})
     {\widetilde Z}^{\star}_{b_1} (x^+;\mathbf{p}_{1}){\widetilde Z}^{\star}_{b_2} (x^+;\mathbf{p}_{2}) \\ + \int\!d^{3}\mathbf{p}_{1}\, d^{3}\mathbf{p}_{2}\, {\widetilde \Omega}_{2}^{a b_1 \left \{b_2  \right \} }(\mathbf{P}; \mathbf{p}_{1} ,\left \{ \mathbf{p}_{2}  \right \}) {\widetilde Z}^{\star}_{b_1} (x^+;\mathbf{p}_{1}) {\widetilde Z}^{\bullet}_{b_2} (x^+;\mathbf{p}_{2}) \, .
    \label{eq:astar_2}
\end{multline}
The above  kernels represent the momentum space version of $\Xi_{1,1}^{ab_1 b_{2}}(\mathbf{x};\mathbf{y}_1, \mathbf{y}_{2})$,  $\Xi_{2,0}^{ab_1 b_{2}}(\mathbf{x};\mathbf{y}_1, \mathbf{y}_{2})$, $\Lambda_{2,0}^{ab_1 b_{2}}(\mathbf{x};\mathbf{y}_1, \mathbf{y}_{2})$ and $\Lambda_{1,1}^{ab_1 b_{2}}(\mathbf{x};\mathbf{y}_1, \mathbf{y}_{2})$ respectively, introduced in Eqs.~\eqref{eq:Abullet_to_Z}-\eqref{eq:Astar_to_Z}.
All the above relations are three dimensional Fourier transforms performed for a fixed light-come time $x^+$. Since the kernels are independent of the minus component of momentum $k^-$, the four dimensional transforms will only account for an extra delta function for the minus component conservation.

In momentum space, the kinetic and triple-gluon terms of the Yang-Mills action \eqref{eq:YM_LC_action} read
\begin{equation}
     \mathcal{L}_{+-}^{\left(\mathrm{LC}\right)}=\int d^{4}{p}_{1}d^{4}{p}_{2}\,\delta^{4}\left({p}_{1}+{p}_{2}\right)\, {p}_{1}^{2}\,\,
\widetilde{A}_{a}^{\bullet}\left({p}_{1}\right)\widetilde{A}_{a}^{\star}\left({p}_{2}\right)\,,
\label{eq:kin_sym}
\end{equation}
\begin{multline}
    \mathcal{L}_{++-}^{\left(\mathrm{LC}\right)}=\int d^{3}\mathbf{p}_{1}d^{3}\mathbf{p}_{2}d^{3}\mathbf{p}_{3}\,\delta^{3}\left(\mathbf{p}_{1}+\mathbf{p}_{2}+\mathbf{p}_{3}\right)\widetilde{V}_{++-}^{abc}\left(\mathbf{p}_{1},\mathbf{p}_{2},\mathbf{p}_{3}\right)\,\\
\widetilde{A}_{a}^{\bullet}\left(x^+;\mathbf{p}_{1}\right)\widetilde{A}_{b}^{\bullet}\left(x^+;\mathbf{p}_{2}\right)\widetilde{A}_{c}^{\star}\left(x^+;\mathbf{p}_{3}\right)\,,
\label{eq:3g++-}
\end{multline}
with the helicity triple-gluon vertex
\begin{equation}
\widetilde{V}_{++-}^{abc}\left(\mathbf{p}_{1},\mathbf{p}_{2},\mathbf{p}_{3}\right)=-igf^{abc}\left(\frac{p_{1}^{\star}}{p_{1}^{+}}-\frac{p_{2}^{\star}}{p_{2}^{+}}\right)p_{3}^{+}\,
\label{eq:3g++-V}
\end{equation}
and
\begin{multline}
    \mathcal{L}_{--+}^{\left(\mathrm{LC}\right)}=\int d^{3}\mathbf{p}_{1}d^{3}\mathbf{p}_{2}d^{3}\mathbf{p}_{3}\,\delta^{3}\left(\mathbf{p}_{1}+\mathbf{p}_{2}+\mathbf{p}_{3}\right)\widetilde{V}_{--+}^{abc}\left(\mathbf{p}_{1},\mathbf{p}_{2},\mathbf{p}_{3}\right)\, \\
    \widetilde{A}_{a}^{\star}\left(x^+;\mathbf{p}_{1}\right)\widetilde{A}_{b}^{\star}\left(x^+;\mathbf{p}_{2}\right)\widetilde{A}_{c}^{\bullet}\left(x^+;\mathbf{p}_{3}\right)\,,
\label{eq:3g+--}
\end{multline}
with
\begin{equation}
\widetilde{V}_{--+}^{abc}\left(\mathbf{p}_{1},\mathbf{p}_{2},\mathbf{p}_{3}\right)=-igf^{abc}\left(\frac{p_{1}^{\bullet}}{p_{1}^{+}}-\frac{p_{2}^{\bullet}}{p_{2}^{+}}\right)p_{3}^{+}\,.\label{eq:3g+--V}
\end{equation}
Above, we have used the 3-dimensional Fourier transform at a constant light-cone time $x^+$ for the interaction terms, and the full 4 dimensional Fourier transform for the kinetic term, because the inverse propagator acts also on the light-cone time (i.e. it contains $\partial_+$ operator).
Since the triple-gluon vertices are independent of $x^+$, we can directly substitute the $A$ fields in term of $Z$ fields, at a constant $x^+$ time.

In order to deal with the kinetic term, we need (\ref{eq:A_bull_2}) and (\ref{eq:astar_2}) written fully in momentum space. As mentioned, this is straightforward and accounts for a minus momentum component conservation delta in the kernels. 
We have, up to the third order in $Z$ fields,
\begin{multline}
    \mathcal{L}_{+-}=\int d^{4}{p}_{1}d^{4}{p}_{2}\,\delta^{4}\left({p}_{1}+{p}_{2}\right)\, {p}_{1}^{2}\,\,
\Bigg\{\Bigg[\int\!d^{4}{q}_{1} d^{4}{q}_{2}\,\, \overline{\widetilde \Omega}\,^{a c_1 \left \{c_2 \right \} }_2({p}_{1}; {q}_{1} ,\left \{ {q}_{2} \right \}) \\
\times {\widetilde Z}^{\bullet}_{c_1} ({q}_{1}) {\widetilde Z}^{\star}_{c_2} ({q}_{2})     +  {\widetilde \Psi}_{2}^{a \left \{c_1 c_2 \right \} }({p}_{1}; \left \{{q}_{1}, {q}_{2} \right \}) {\widetilde Z}^{\bullet}_{c_1} ({q}_{1}){\widetilde Z}^{\bullet}_{c_2} ({q}_{2}) \,\Bigg]\widetilde{Z}_{a}^{\star}\left({p}_{2}\right)\, \\ +\widetilde{Z}_{a}^{\bullet}\left({p}_{1}\right) 
\times \Bigg[ \int\!d^{4}{q}_{1}\, d^{4}{q}_{2}\,\, \overline{\Psi}\,^{a \left \{c_1 c_2 \right \} }_2({p}_{2}; \left \{{q}_{1},   {q}_{2} \right \})      {\widetilde Z}^{\star}_{c_1} ({q}_{1}){\widetilde Z}^{\star}_{c_2} ({q}_{2})\\
     +  {\widetilde \Omega}_{2}^{a c_1 \left \{c_2  \right \} }({p}_{2}; {q}_{1} ,\left \{ {q}_{2}  \right \}) {\widetilde Z}^{\star}_{c_1} ({q}_{1}) {\widetilde Z}^{\bullet}_{c_2} ({q}_{2})\Bigg]\Bigg\} \,.
     \label{eq:kin_subs}
\end{multline}
 Let us first consider the terms that have (${\widetilde{Z}^\star}{\widetilde{Z}^\star}{\widetilde{Z}^\bullet}$) field configuration:  
\begin{multline}
    \mathcal{T}_{--+}=\int d^{4}{p}_{1}d^{4}{p}_{2}\,\delta^{4}\left({p}_{1}+{p}_{2}\right)\, {p}_{1}^{2}\,\, \Bigg[\int\!d^{4}{q}_{1} d^{4}{q}_{2} \,\, \overline{\widetilde \Omega}\,^{a c_1 \left \{c_2 \right \} }_2({p}_{1}; {q}_{1} ,\left \{ {q}_{2} \right \})\\
\times {\widetilde Z}^{\bullet}_{c_1} ({q}_{1}) {\widetilde Z}^{\star}_{c_2} ({q}_{2}) \widetilde{Z}_{a}^{\star}\left({p}_{2}\right)\,  +\widetilde{Z}_{a}^{\bullet}\left({p}_{1}\right)
    \int\!d^{4}{q}_{1}\, d^{4}{q}_{2}\,\, \overline{\widetilde{\Psi}}\,^{a \left \{c_1 c_2 \right \} }_2({p}_{2}; \left \{{q}_{1},   {q}_{2} \right \}) {\widetilde Z}^{\star}_{c_1} ({q}_{1}){\widetilde Z}^{\star}_{c_2} ({q}_{2}) \Bigg] \, .
     \label{eq:v--+_tr}
\end{multline}
Integrating the first  and second term over $p_1$ and $p_2$ respectively, we see that each term will have only three momentum variables. Both terms can be combined into one integral by renaming the momentum variables to $p_1, p_2, p_3$, and color indices to $b_1$, $b_2$, $b_3$. With this we have  
\begin{multline}
   \mathcal{T}_{--+}=\int d^{4}{p}_{1}d^{4}{p}_{2}d^{4}{p}_{3}\, \Bigg[ {p}_{2}^{2}\,\,\overline{\widetilde \Omega}\,^{b_2 b_3 \left \{b_1 \right \} }_2(-{p}_{2}; {p}_{3} ,\left \{ {p}_{1} \right \}) \\ +  {p}_{3}^{2}\overline{\widetilde{\Psi}}\,^{b_3 \left \{b_1 b_2 \right \} }_2(-{p}_{3}; \left \{{p}_{1},   {p}_{2} \right \}) \Bigg]{\widetilde Z}^{\star}_{b_1} ({p}_{1}){\widetilde Z}^{\star}_{b_2} ({p}_{2})\widetilde{Z}_{b_3}^{\bullet}\left({p}_{3}\right) \, .
   \label{eq:v--t_com}
\end{multline}
In order to bring the above expression to the constant light cone time $x^+$ we introduce the following 
auxiliary
fields (both for $\widetilde{Z}^\bullet$ and $\widetilde{Z}^\star$) 
\begin{equation}
    {\widetilde Z}_{b_i}(p_i) = \frac{{\widetilde \kappa}_{b_i}(p_i) }{p_i^2} \, .
    \label{eq:kappa}
\end{equation}
Substituting (\ref{eq:kappa}) in (\ref{eq:v--t_com}) we obtain
\begin{multline}
   \mathcal{T}_{--+}=\int d^{4}{p}_{1}d^{4}{p}_{2}d^{4}{p}_{3}\, \Bigg[ {p}_{2}^{2}\,\,\overline{\widetilde \Omega}\,^{b_2 b_3 \left \{b_1 \right \} }_2(-{p}_{2}; {p}_{3} ,\left \{ {p}_{1} \right \}) \\ +  {p}_{3}^{2}\overline{\widetilde{\Psi}}\,^{b_3 \left \{b_1 b_2 \right \} }_2(-{p}_{3}; \left \{{p}_{1},   {p}_{2} \right \}) \Bigg]\frac{{\widetilde \kappa}^{\star}_{b_1}(p_1) }{p_1^2}\frac{{\widetilde \kappa}^{\star}_{b_2}(p_2) }{p_2^2}\frac{{\widetilde \kappa}^{\bullet}_{b_3}(p_3) }{p_3^2} \, .
   \label{eq:v--t_kap}
\end{multline}
Next, we integrate out the minus momentum components. To this end, the first term in (\ref{eq:v--t_kap}) can be rewritten as
\begin{multline}
    \int d^{3}\mathbf{p}_{1}d^{3}\mathbf{p}_{2}d^{3}\mathbf{p}_{3}\,\,\int dp_1^-\, dp_2^-\ dp_3^- dP^-\, \int dz^+ \int dy^+
    e^{iz^+(P^- - p_2^-)} \,\,    e^{iy^+(P^- +p_1^- + p_3^-)}    (-p_{13}^+)(P^- + \hat{p}_{13})\\
    \overline{\widetilde \Omega}\,^{b_2 b_3 \left \{b_1 \right \} }_2(-\mathbf{p}_2; \mathbf{p}_{3} ,\left \{ \mathbf{p}_{1} \right \}) \frac{1}{2p_1^+ [p_1^- - \hat{p}_{1} + i\epsilon]}\frac{1}{2p_2^+ [p_2^- - \hat{p}_{2} + i\epsilon]}\\
    \frac{1}{2p_3^+ [p_3^- - \hat{p}_{3} + i\epsilon]}{\widetilde \kappa}^{\star}_{b_1}(p_1){\widetilde \kappa}^{\star}_{b_2}(p_2){\widetilde \kappa}^{\star}_{b_3}(p_3) \, .
    \label{eq:v--t1}
\end{multline}
where we introduced the notation
\begin{equation}
    \hat{p} = \frac{p^\bullet p^\star}{p^+} \, ,
    \label{eq:hat_def}
\end{equation}
for any momentum $p$.
This leads to 
\begin{multline}
    \int d^{3}\mathbf{p}_{1}d^{3}\mathbf{p}_{2}d^{3}\mathbf{p}_{3}\,\,\,p_{13}^+(\hat{p}_{1} + \hat{p}_{3} -\hat{p}_{13})\,\,
    \overline{\widetilde \Omega}\,^{b_2 b_3 \left \{b_1 \right \} }_2(-\mathbf{p}_2; \mathbf{p}_{3} ,\left \{ \mathbf{p}_{1} \right \}) \frac{{(i\pi)}^3}{p_1^+ p_2^+ p_3^+} \\ \times \Bigg[ \prod_{i=1}^{3}\Theta (-p_i^+) \frac{i}{\hat{p}_{1}+\hat{p}_{2} + \hat{p}_{3} + i\epsilon}
    +\prod_{i=1}^{3}\Theta (p_i^+) {(-1)}^3 \frac{-i}{\hat{p}_{1}+\hat{p}_{2} + \hat{p}_{3} + i\epsilon} \Bigg]\\
    \times {\widetilde \kappa}^{\star}_{b_1}\left(\hat{p}_{1};\mathbf{p}_{1}\right){\widetilde \kappa}^{\star}_{b_2}\left(\hat{p}_{2};\mathbf{p}_{2}\right){\widetilde \kappa}^{\star}_{b_3}\left(\hat{p}_{3};\mathbf{p}_{3}\right) \, ,
    \label{eq:v--+1_cl}
    \end{multline}
where $\Theta (p_i^+)$ is Heaviside step function. This may be rewritten as
\begin{multline}
  \int dx^+\, \int d^{3}\mathbf{p}_{1}d^{3}\mathbf{p}_{2}d^{3}\mathbf{p}_{3}\,\,\,p_{13}^+(\hat{p}_{1} + \hat{p}_{3} -\hat{p}_{13})\,\,
  \overline{\widetilde \Omega}\,^{b_2 b_3 \left \{b_1 \right \} }_2(-\mathbf{p}_2; \mathbf{p}_{3} ,\left \{ \mathbf{p}_{1} \right \}) \\ 
  \times {\widetilde Z}^{\star}_{b_1} (x^+;\mathbf{p}_{1}){\widetilde Z}^{\star}_{b_2} (x^+;\mathbf{p}_{2})\widetilde{Z}_{b_3}^{\bullet}\left(x^+;\mathbf{p}_{3}\right) \, ,
  \label{eq:v--+vcl}
\end{multline}
where, in going from (\ref{eq:v--+1_cl}) to (\ref{eq:v--+vcl}), we used the following relation
\begin{multline}
     \int dx^+\, \int d^{3}\mathbf{p}_{1}\cdots d^{3}\mathbf{p}_{n}\,\,\,
  {\widetilde f}(\mathbf{p}_1 \cdots \mathbf{p}_{n})
   {\widetilde Z}^{\star}_{b_1} (x^+;\mathbf{p}_{1}) \cdots \widetilde{Z}_{b_n}^{\bullet}\left(x^+;\mathbf{p}_{n}\right)\\ = 
  \int d^{3}\mathbf{p}_{1}\cdots d^{3}\mathbf{p}_{n}\,\,\, {\widetilde f}(\mathbf{p}_1 \cdots \mathbf{p}_{n})\frac{{(i\pi)}^n}{p_1^+ \cdots p_n^+} \Bigg[ \prod_{i=1}^{n}\Theta (-p_i^+) \frac{i}{\hat{p}_{1}+\cdots + \hat{p}_{n} + i\epsilon} \\
    +\prod_{i=1}^{n}\Theta (p_i^+) {(-1)}^n \frac{-i}{\hat{p_{1}}+\cdots + \hat{p_{n}} + i\epsilon} \Bigg]{\widetilde \kappa}^{\star}_{b_1}\left(\hat{p}_{1};\mathbf{p}_{1}\right) \cdots {\widetilde \kappa}^{\star}_{b_n}\left(\hat{p}_{n};\mathbf{p}_{n}\right) \, .
\end{multline}
Above, ${\widetilde f}(\mathbf{p}_1 \cdots \mathbf{p}_{n})$ represents any generic function not depending on the minus momentum components (or the light-cone time).
In a similar way, the second term in (\ref{eq:v--t_kap}) gives
\begin{multline}
  \int dx^+\, \int d^{3}\mathbf{p}_{1}d^{3}\mathbf{p}_{2}d^{3}\mathbf{p}_{3}\,\,\,p_{12}^+(\hat{p}_{1} + \hat{p}_{2} -\hat{p}_{12})
  \overline{\widetilde{\Psi}}\,^{b_3 \left \{b_1 b_2 \right \} }_2(-\mathbf{p}_3; \left \{\mathbf{p}_{1},   \mathbf{p}_{2} \right \}) \\
  \times {\widetilde Z}^{\star}_{b_1} (x^+;\mathbf{p}_{1}) {\widetilde Z}^{\star}_{b_2} (x^+;\mathbf{p}_{2})\widetilde{Z}_{b_3}^{\bullet}\left(x^+;\mathbf{p}_{3}\right) \, .
  \label{eq:v--+vcl2}
\end{multline}
Combining Eq.~(\ref{eq:v--+vcl}) and (\ref{eq:v--+vcl2}) we get
\begin{multline}
    \mathcal{T}_{--+}=\int dx^+\, \int d^{3}\mathbf{p}_{1}d^{3}\mathbf{p}_{2}d^{3}\mathbf{p}_{3}\,
    \Bigg[p_{13}^+(\hat{p}_{1} + \hat{p}_{3} -\hat{p}_{13}) 
  \,\, \overline{\widetilde \Omega}\,^{b_2 b_3 \left \{b_1 \right \} }_2(-\mathbf{p}_2; \mathbf{p}_{3} ,\left \{ \mathbf{p}_{1} \right \}) \\ + p_{12}^+(\hat{p}_{1} + \hat{p}_{2} -\hat{p}_{12})
  \,\, \overline{\widetilde{\Psi}}\,^{b_3 \left \{b_1 b_2 \right \} }_2(-\mathbf{p}_3; \left \{\mathbf{p}_{1},   \mathbf{p}_{2} \right \})\Bigg] {\widetilde Z}^{\star}_{b_1} (x^+;\mathbf{p}_{1}){\widetilde Z}^{\star}_{b_2} (x^+;\mathbf{p}_{2})\widetilde{Z}_{b_3}^{\bullet}\left(x^+;\mathbf{p}_{3}\right) \, .
  \label{eq:v--t_can}
\end{multline}
Using $p_{ij}^+(\hat{p_{i}} + \hat{p_{j}} -\hat{p_{ij}}) =-{\widetilde v}_{(i)(j)}{\widetilde v}^{\ast}_{(j)(i)}$ and substituting for $\overline{\widetilde \Omega}$ and $\overline{\widetilde{\Psi}}$ the kernels from \eqref{eq:omegaBar_kernel} and \eqref{eq:psiBar_kernel}, respectively, after a bit of algebra we obtain
\begin{multline}
    \mathcal{T}_{--+}=\int dx^+\, \int d^{3}\mathbf{p}_{1}d^{3}\mathbf{p}_{2}d^{3}\mathbf{p}_{3}\,\delta^{3}\left(\mathbf{p}_{1}+\mathbf{p}_{2}+\mathbf{p}_{3}\right) \\
    \times \Big( ig f^{b_1 b_2 b_3} p_3^+ {v}^{\ast}_{12}  \Big) 
  \times {\widetilde Z}^{\star}_{b_1} (x^+;\mathbf{p}_{1}){\widetilde Z}^{\star}_{b_2} (x^+;\mathbf{p}_{2})\widetilde{Z}_{b_3}^{\bullet}\left(x^+;\mathbf{p}_{3}\right) \, ,
  \label{eq:v--t_can1}
\end{multline}
where $\widetilde{v}_{ij}=p_i^+v_{ji}$. Comparing this with (\ref{eq:3g+--V}), we may rewrite the above as
\begin{multline}
    \mathcal{T}_{--+}=\int dx^+\, \int d^{3}\mathbf{p}_{1}d^{3}\mathbf{p}_{2}d^{3}\mathbf{p}_{3}\,\delta^{3}\left(\mathbf{p}_{1}+\mathbf{p}_{2}+\mathbf{p}_{3}\right) \\
    \times \Big( - \widetilde{V}_{--+}^{b_1 b_2 b_3}\left(\mathbf{p}_{1},\mathbf{p}_{2},\mathbf{p}_{3}\right) \Big) 
  \times {\widetilde Z}^{\star}_{b_1} (x^+;\mathbf{p}_{1}){\widetilde Z}^{\star}_{b_2} (x^+;\mathbf{p}_{2})\widetilde{Z}_{b_3}^{\bullet}\left(x^+;\mathbf{p}_{3}\right) \, .
  \label{eq:v--t_comp}
\end{multline}
This cancels out the triple-gluon vertex coming from the Yang-Mills action \eqref{eq:3g+--}  when we substitute the first order expansion of $\widetilde{A}^\bullet$ and $\widetilde{A}^\star$ in terms of $\widetilde{Z}^\bullet$ and $\widetilde{Z}^\star$ fields. 

In exactly same fashion, the cancellation of the other triple-gluon vertex $\widetilde{V}_{++-}^{b_1 b_2 b_3}\left(\mathbf{p}_{1},\mathbf{p}_{2},\mathbf{p}_{3}\right)$ can be shown.
%-----------------------------------------------
%-----------------------------------------------
\section{Six point $\overline{\mathrm{MHV}}$ amplitude.}
\label{sec:AppB}

\begin{figure}[h]
    \centering
 \includegraphics[width=13.5cm]{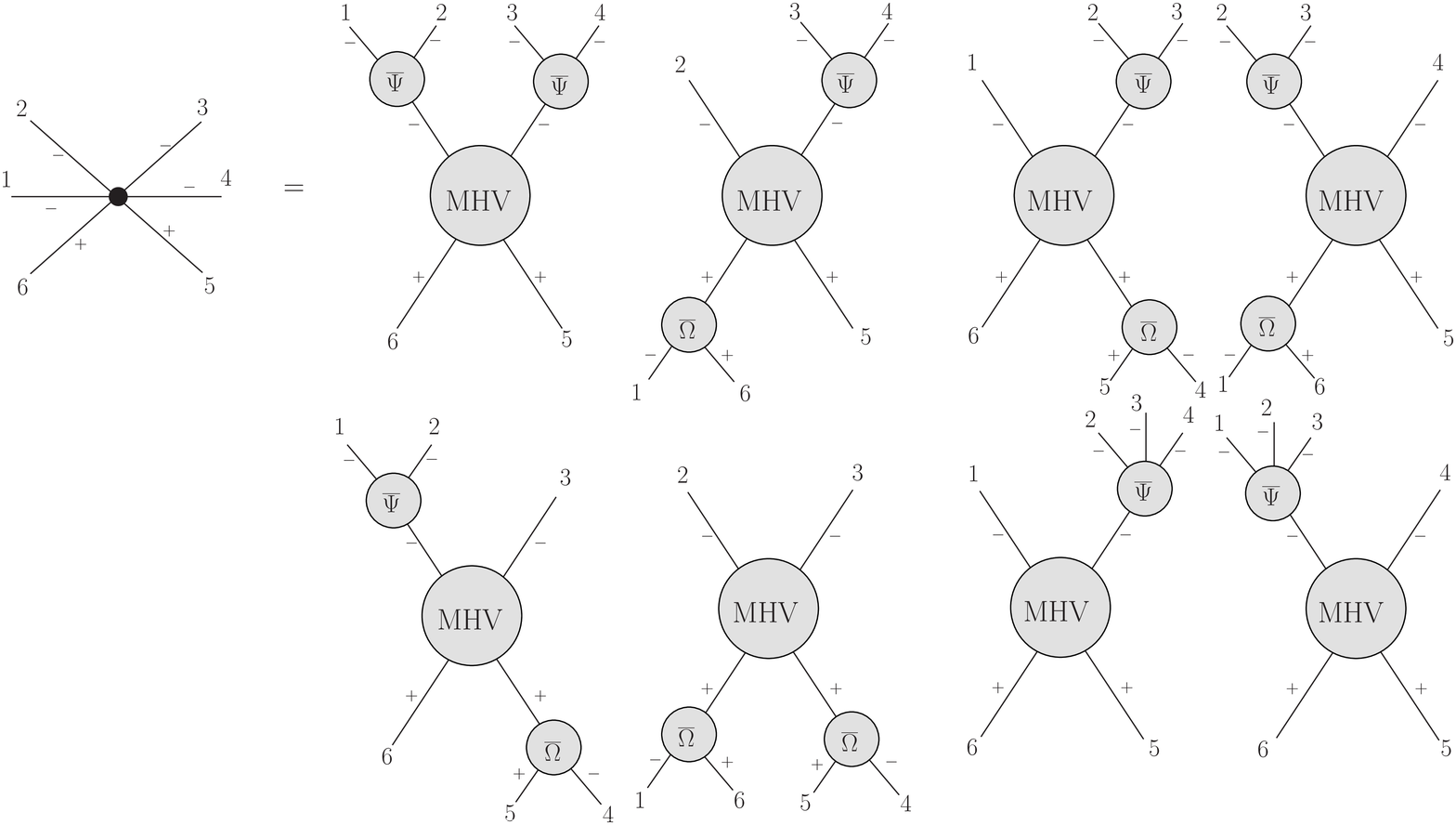}\\
 \includegraphics[width=5.5cm]{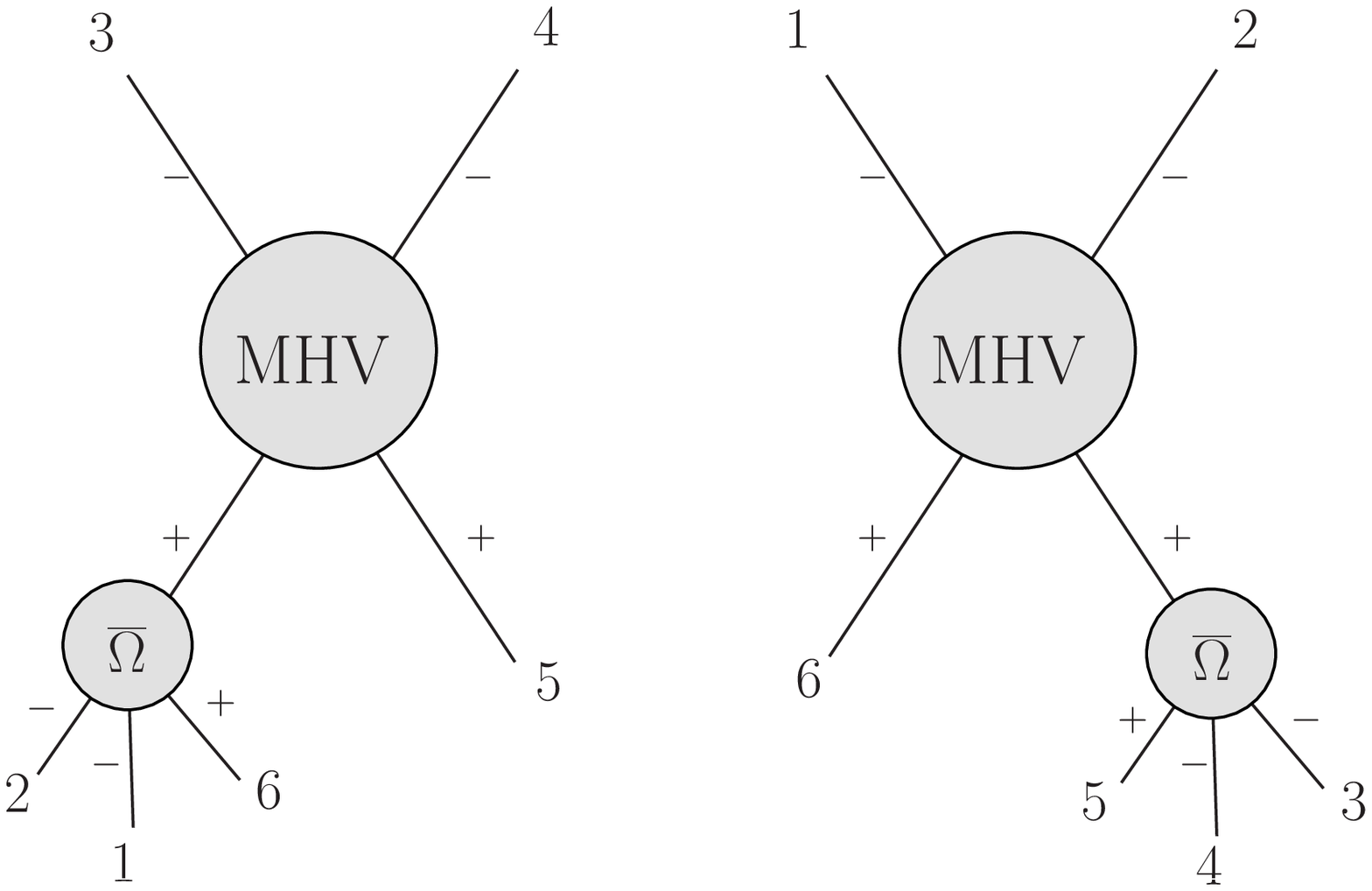}
    \caption{\small 
    The contributions to the color-ordered 6 point $\overline{\text{MHV}}$ vertex, with helicity $(----++)$.
    }
    \label{fig:6gmhv_b}
\end{figure}

In this appendix we show the details associated with the calculation of the 6-point $\overline{\mathrm{MHV}}$ $(----++)$ amplitude. As mentioned previously, the $\overline{\text{MHV}}$ amplitudes are always given by a single vertex in the action \eqref{eq:Z_action1}.  For the color ordered amplitude we need the  $\mathcal{U}(1^-,2^-,3^-,4^-,5^+,6^+)$ vertex which is given by the formula \eqref{eq:Z_gen_ker}. In Fig.~\ref{fig:6gmhv_b} we show all the contributing terms. Using the explicit expressions we get:

\begin{multline}
 \mathcal{U}(1^-,2^-,3^-,4^-,5^+,6^+) = g^4
  \Bigg[  \Bigg( \left(\frac{p_{12} ^{+}}{p_{34}^{+}}\right)^{2}
\frac{\widetilde{v}_{(34)({12})}^{*4}}{\widetilde{v}_{({12}){6}}^{*}\widetilde{v}_{{6}5}^{*}\widetilde{v}_{{5}(34)}^{*}  \widetilde{v}_{({34})({12})}^{*} } \times  \frac{{\widetilde v}_{({12})1}}{{\widetilde v}_{21}{\widetilde v}_{1({12})}}\times  \frac{{\widetilde v}_{({34})3}}{{\widetilde v}_{43}{\widetilde v}_{3({34})}} \Bigg) \\
+ \Bigg(  \left(\frac{p_{2} ^{+}}{p_{34}^{+}}\right)^{2}
\frac{\widetilde{v}_{(34){2}}^{*4}}{\widetilde{v}_{2({16})}^{*}\widetilde{v}_{{(16)5}}^{*}\widetilde{v}_{{5}(34)}^{*}  \widetilde{v}_{({34}){2}}^{*} } \times \frac{{\widetilde v}_{({34})3}}{{\widetilde v}_{43}{\widetilde v}_{3({34})}} \times \left(\frac{p_{6} ^{+}}{p_{16}^{+}}\right)^{2} \frac{{\widetilde v}_{({16})6}}{{\widetilde v}_{16}{\widetilde v}_{6({16})}}   \Bigg)\\
+ \Bigg(  \left(\frac{p_{1} ^{+}}{p_{23}^{+}}\right)^{2}
\frac{\widetilde{v}_{(23){1}}^{*4}}{\widetilde{v}_{{16}}^{*}\widetilde{v}_{{6}\left(45\right)}^{*}\widetilde{v}_{({45)}(23)}^{*}  \widetilde{v}_{({23}){1}}^{*} } \times \frac{{\widetilde v}_{({23})2}}{{\widetilde v}_{32}{\widetilde v}_{2({23})}} \times \left(\frac{p_{5} ^{+}}{p_{45}^{+}}\right)^{2} \frac{{\widetilde v}_{({45})4}}{{\widetilde v}_{54}{\widetilde v}_{4({45})}}   \Bigg)\\
+ \Bigg( \left(\frac{p_{23} ^{+}}{p_{4}^{+}}\right)^{2}
\frac{\widetilde{v}_{4({23})}^{*4}}{\widetilde{v}_{({23})({16})}^{*}\widetilde{v}_{({16})5}^{*}\widetilde{v}_{{54}}^{*}  \widetilde{v}_{{4}({23})}^{*} } \times\frac{{\widetilde v}_{({23})2}}{{\widetilde v}_{32}{\widetilde v}_{2({23})}}  \times \left(\frac{p_{6} ^{+}}{p_{16}^{+}}\right)^{2} \frac{{\widetilde v}_{({16})6}}{{\widetilde v}_{16}{\widetilde v}_{6({16})}}   \Bigg) \\
+ \Bigg(  \left(\frac{p_{12} ^{+}}{p_{3}^{+}}\right)^{2}
\frac{\widetilde{v}_{3({12})}^{*4}}{\widetilde{v}_{({12}){6}}^{*}\widetilde{v}_{{6}\left(45\right)}^{*}\widetilde{v}_{({45)}3}^{*}  \widetilde{v}_{{3}({12})}^{*} } \times  \left(\frac{p_{5} ^{+}}{p_{45}^{+}}\right)^{2} \frac{{\widetilde v}_{({45})4}}{{\widetilde v}_{54}{\widetilde v}_{4({45})}} \times \frac{{\widetilde v}_{({12})1}}{{\widetilde v}_{21}{\widetilde v}_{1({12})}}  \Bigg) \\
+ \Bigg(  \left(\frac{p_{2} ^{+}}{p_{3}^{+}}\right)^{2}
\frac{\widetilde{v}_{32}^{*4}}{\widetilde{v}_{2({16})}^{*}\widetilde{v}_{{(16)(45)}}^{*}\widetilde{v}_{(45)3}^{*}  \widetilde{v}_{{32}}^{*} } \left(\frac{p_{5} ^{+}}{p_{45}^{+}}\right)^{2} \frac{{\widetilde v}_{({45})4}}{{\widetilde v}_{54}{\widetilde v}_{4({45})}} \left(\frac{p_{6} ^{+}}{p_{16}^{+}}\right)^{2} \frac{{\widetilde v}_{({16})6}}{{\widetilde v}_{16}{\widetilde v}_{6({16})}}   \Bigg)\\
- \Bigg(  \left(\frac{p_{1} ^{+}}{p_{234}^{+}}\right)^{2}
\frac{\widetilde{v}_{(234){1}}^{*4}}{\widetilde{v}_{{1}{6}}^{*}\widetilde{v}_{{65}}^{*}\widetilde{v}_{{5}(234)}^{*}  \widetilde{v}_{({234}){1}}^{*} } \times  \frac{{\widetilde v}_{({234})2}}{{\widetilde v}_{43}{\widetilde v}_{32}{\widetilde v}_{2({234})}}  \Bigg) \\
- \Bigg( \left(\frac{p_{123} ^{+}}{p_{4}^{+}}\right)^{2}
\frac{\widetilde{v}_{4({123})}^{*4}}{\widetilde{v}_{({123}){6}}^{*}\widetilde{v}_{{65}}^{*}\widetilde{v}_{{54}}^{*}  \widetilde{v}_{{4}({123})}^{*} } \times  \frac{{\widetilde v}_{({123})1}}{{\widetilde v}_{32}{\widetilde v}_{21}{\widetilde v}_{1({123})}} \Bigg) \\
- \Bigg(  \left(\frac{p_{3} ^{+}}{p_{4}^{+}}\right)^{2}
\frac{\widetilde{v}_{43}^{*4}}{\widetilde{v}_{3{(612)}}^{*}\widetilde{v}_{{(612)5}}^{*}\widetilde{v}_{54}^{*}  \widetilde{v}_{43}^{*} } \times \left(\frac{p_{6} ^{+}}{p_{612}^{+}}\right)^{2} \frac{{\widetilde v}_{({612})6}}{{\widetilde v}_{21}{\widetilde v}_{16}{\widetilde v}_{6({612})}}   \Bigg)\\
- \Bigg( \left(\frac{p_{1} ^{+}}{p_{2}^{+}}\right)^{2}
\frac{\widetilde{v}_{21}^{*4}}{\widetilde{v}_{{1}{6}}^{*}\widetilde{v}_{{6}\left(345\right)}^{*}\widetilde{v}_{({345)}2}^{*}  \widetilde{v}_{21}^{*} } \times \left(\frac{p_{5} ^{+}}{p_{345}^{+}}\right)^{2} \frac{{\widetilde v}_{({345})3}}{{\widetilde v}_{54}{\widetilde v}_{43}{\widetilde v}_{3({345})}}   \Bigg) \Bigg] \, .
 \label{eq:6g_MHVbar_Z}
\end{multline}

We checked, that the above expression reduces in the on-shell limit to the known expression:
\begin{equation}
    \mathcal{A}(1^-,2^-,3^-,4^-,5^+,6^+) =
 g^4 \left(\frac{p_{5} ^{+}}{p_{6}^{+}}\right)^{2}
\frac{\widetilde{v}_{65}^{4}}{\widetilde{v}_{16}\widetilde{v}_{65}\widetilde{v}_{54}  \widetilde{v}_{43}\widetilde{v}_{32}\widetilde{v}_{21} } \, .
 \label{eq:6G_MHVbar_onshell}
\end{equation}

%--------------------------------------------------------------------------------

%-----------------------------------------------------------------------
\end{document}